%% file: ms1024.tex
\documentclass{aa}
\usepackage{graphicx, natbib}
\bibpunct[]{(}{)}{;}{a}{}{,}

\begin{document}

\title{The impact of bars on the mid-infrared dust emission of spiral
       galaxies: global and circumnuclear properties\thanks{Based on
       observations with ISO, an ESA project with instruments funded by
       ESA Member States (especially the PI countries: France, Germany,
       the Netherlands and the United Kingdom) and with the participation
       of ISAS and NASA.}}

\author{H. Roussel\inst{1}
\and M. Sauvage\inst{1}
\and L. Vigroux\inst{1}
\and A. Bosma\inst{2}
\and C. Bonoli\inst{3}
\and \\
     P. Gallais\inst{1}
\and T. Hawarden\inst{4}
\and S. Madden\inst{1}
\and P. Mazzei\inst{3}}

\institute{DAPNIA/Service d'Astrophysique, CEA/Saclay, 91191 Gif-sur-Yvette cedex, France
\and Observatoire de Marseille, 2 Place Le Verrier, 13248 Marseille cedex 4, France
\and Osservatorio Astronomico di Padova, 5 Vicolo dell'Osservatorio, 35122 Padova, Italy
\and Joint Astronomy Center, 660 N. A'ohoku Place, Hilo, Hawaii 96720, USA}

\titlerunning{Impact of bars on MIR dust emission of spirals}

\offprints{H. Roussel (e-mail: hroussel@cea.fr)}

\date{Received 14 December 2000 / Accepted 22 March 2001}

\maketitle

\begin{abstract}
We study the mid-infrared properties of a sample of 69 nearby spiral
galaxies, selected to avoid Seyfert activity contributing a significant
fraction of the central energetics, or strong tidal interaction,
and to have normal infrared luminosities. These observations were
obtained with ISOCAM, which provides an angular resolution of the order
of $10\arcsec$ (half-power diameter of the point spread function) and
low-resolution spectro-imaging information. Between
5 and 18\,$\mu$m, we mainly observe two dust phases, aromatic infrared
bands and very small grains, both out of thermal equilibrium. On this
sample, we show that the global $F_{15}/F_7$ colors of galaxies are very
uniform, the only increase being found in early-type strongly barred
galaxies, consistent with previous IRAS studies. The $F_{15}/F_7$ excesses
are unambiguously due to galactic central regions where bar-induced
starbursts occur. However, the existence of strongly barred early-type
galaxies with normal circumnuclear colors indicates that the relationship
between a distortion of the gravitational potential and a central
starburst is not straightforward. \\
As the physical processes at work in central regions are in principle
identical in barred and unbarred galaxies, and since this is where the
mid-infrared activity is mainly located, we investigate the mid-infrared
circumnuclear properties of all the galaxies in our sample. We show how
surface brightnesses and colors are related to both the available molecular
gas content and the mean age of stellar populations contributing to dust
heating. Therefore, the star formation history in galactic central regions
can be constrained by their position in a color-surface brightness
mid-infrared diagram.
\keywords{galaxies: spiral -- galaxies: barred -- galaxies: ISM --
       stars: formation -- infrared: ISM: continuum --
	 infrared: ISM: lines and bands}
\end{abstract}

\section{Introduction}

As the high frequency of bars in galaxies becomes more evident
\citep[{\it e.g.} ][]{Eskridge}, and as new techniques emerge to both
observationally quantify their strength \citep{Seigar, Buta}
and numerically simulate them, their effects on their host galaxies
are of major interest, and in particular, it is worth checking whether
they are indeed very efficient systems to drive nuclear starbursts in
spiral galaxies.

Numerous studies have dealt with the respective star formation
properties of barred and non-barred spirals, mostly in the infrared,
since this is the wavelength regime where starbursts are expected to be
most easily detectable. Yet conclusions derived from such studies
appear to contradict each other, partly because the different
selection criteria result in samples with a more or less pronounced
bias toward starburst objects. For instance, in the IR-bright sample
analyzed by \citet{Hawarden}, an important fraction of SB and SAB
galaxies \citep[respectively strongly barred and weakly barred spirals in
the classification of][]{Vaucouleurs} shows a 25\,$\mu$m emission
excess (with respect to 12 and 100\,$\mu$m) absent in the SA subsample
(non-barred spirals), which can be accounted for by a highly increased
contribution of Galactic-like H{\scriptsize II} regions to the total emission.
On the other hand, \citet{Isobe}, using a volume-limited sample and
performing a survival analysis to take into account the frequent IRAS
non-detections, found that the far-IR to blue flux ratio ($F_{\rm FIR}/F_{\rm B}$)
is rather independent of the bar class. The contradiction is marginal
since $F_{\rm FIR}/F_{\rm B}$ does not give a direct estimation of the star
formation activity, especially when dealing with quiescent normal galaxies:
the blue light originates partly from young stars and, as \citet{Isobe}
emphasize, $F_{\rm FIR}/F_{\rm B}$ depends on the amount and spatial distribution
of dust with respect to stars. The relationship between the 25\,$\mu$m excess,
quantified by $F_{25}/F_{12}$, and $F_{\rm FIR}/F_{\rm B}$
in a galaxy sample with good quality data is indeed highly dispersed.
\citet{Huang} investigated the 25\,$\mu$m excess as a function of IR
brightness and reconciled the two previous analyzes: a significant
excess can occur only if $F_{\rm FIR}/F_{\rm B}$ is larger than a threshold
value of $\simeq 0.3$\,. Therefore, a statistical effect of bars on star
formation can be demonstrated only in suitably selected samples.
\citet{Huang} also emphasized that the difference between barred and
unbarred spirals concerns only early types (S0/a to Sbc).

Studies of the infrared excess in barred galaxies mostly rest on the
integrated IRAS measurements, which do not allow the determination of the
nature and location of regions responsible for this excess. However,
dynamical models and observations at other wavelengths give evidence that
the infrared activity should be concentrated in circumnuclear regions
\citep[see for instance the study of NGC\,5383 by][]{Sheth}. In addition,
high-resolution ground-based observations near 10\,$\mu$m of galaxy centers
\citep{Devereux, Telesco} have shown that the dust emission is
more concentrated in barred galaxies.

Theoretically, bars are known to be responsible for large-scale redistribution
of gas through galactic disks. In a strong barred perturbation of the
gravitational potential, shocks develop along the rotation-leading side of
the bar and are associated with strong shear, as shown by \citet{Athanassoulab}
and references therein \citep[also][]{Friedli}.
They induce an increase of gas density which is traced by the thin dust
lanes widely observed in bars, producing a contrasting absorption of optical
light \citep[Prendergast 1962, unpublished;][]{Huntley}. Due to these shocks,
gas loses angular momentum and flows towards the circumnuclear region. This
picture is confirmed by direct observations of inward velocity gradients
across bars in ionized gas lines, CO and H{\scriptsize I}
\citep[{\it e.g.} ][]{Lindblad, Reynaud, Mundell}. \citet{Regan2} derive a
gas accretion rate of $\approx 1$\,M$_{\odot}\,{\rm yr}^{-1}$ into the
circumnuclear ring of NGC\,1530.

Statistical evidence is also found for higher gas concentrations
in the center of barred galaxies
\citep[][, who however observed only SABs, except NGC\,1530]{Sakamoto},
and for more frequent circumnuclear starbursts in barred
galaxies, as reported by \citet{Heckman}, \citet{Hawarden}, \citet{Arsenault}
(who, more exactly, found more probable starbursts in galaxies with both bar and
inner ring, supposed to be a signature of one or two inner Lindblad resonance(s)),
\citet{Huang}, \citet{Martinet} and \citet{Bonatto}.
\citet{Aguerri} has moreover reported that
the global star formation intensity of isolated spirals (mostly of late types)
is correlated with bar strength as quantified by means of its projected axial
ratio, which is surprising in view of the very different timescales of bar
evolution ($\approx 1$\,Gyr) and star formation in kpc-scale regions
($\approx 10^{7-8}$\,yr). Indeed, \citet{Martinet}, using carefully selected
late-type galaxies, found no such correlation, the bar strength being
quantified either by its deprojected axis ratio or its deprojected length
relative to the disk diameter. The fact that only a fraction of strongly barred
galaxies exhibit star formation excess (as evidenced by their IRAS colors)
is explained by these authors with numerical simulations of bar evolution
including gas physics. They show that a strong starburst occurs shortly after
bar formation and quickly fades away (in typically less than 1\,Gyr); meanwhile,
the strength and other properties of the bar evolve, but the bar remains strong
if it was initially strong. The existence of strongly barred galaxies in a
quiescent state is thus to be expected, presumably because the available
gas supply has been consumed in previous bursts.

This paper is aimed at characterizing the mid-infrared excess in barred
galaxies, with the possibility to carry out a detailed and systematic
spatial analysis due to the good angular resolution of ISOCAM (the
half-power beam diameter is less than $10\arcsec$ at 7\,$\mu$m), and
hence to locate unambiguously sites of enhanced infrared activity.
Although dust is a more indirect tracer of young stars than far-ultraviolet
ionizing radiation or optical recombination lines, the infrared 
emission suffers relatively minor extinction effects, which are very
difficult to correct and hamper shorter wavelength studies.
In a companion paper \citep[][, hereafter Paper II]{papier_sfr}, we
have shown that in galactic disks, mid-infrared emission is a reliable
star formation indicator. Here, we concentrate on central regions of
galaxies where the dust heating regime is markedly different from that
in disks.

For this purpose, we have analyzed a sample of 69 nearby spiral
galaxies, imaged at 7 and 15\,$\mu$m with the camera ISOCAM on board ISO
\citep[described by][]{Cesarskyc}. We have also obtained low-resolution
spectroscopic information for a few galaxies, enabling us to identify and
separate the various dust components emitting between 5 and 18\,$\mu$m.
7\,$\mu$m images and $F_{15}/F_7$ flux density ratios of selected regions,
together with optical images, are presented in \citet{atlas} (hereafter
the Atlas). For a description of data reduction and analysis, and a summary
of morphological properties of the sample, the reader is also referred
to the Atlas.

\section{The galaxy sample}
\label{sec:sample}

The sample is intended to be representative of normal quiescent spirals,
and contains galaxies of moderate infrared luminosity. It covers three
guaranteed time programs of ISOCAM. The first one ({\it Cambarre})
consists of nearby barred galaxies, the second one ({\it Camspir})
of a few large-size spirals of special interest (NGC\,1365,
4736, 5194, 5236, 5457 and 6744) and another subsample is drawn from
the Virgo cluster sample of \citet{Boselli} ({\it Virgo} program),
containing relatively fainter and smaller galaxies, both barred and
unbarred. This sample was supplemented by comparable spirals in the
ISOCAM public archive, from the programs {\it Sf\_glx}
\citep{Dale} and {\it Irgal} (PI T. Onaka). All of the observations
were reduced in the same way to form a homogeneous sample. The final
set comprises 69 spiral galaxies at distances between 4 and 60\,Mpc.
We have divided them into three main categories according to
morphological classes in the RC3 \citep{Vaucouleurs}: SBs
(accounting for about half the sample with 37 galaxies), SABs (20
galaxies) and SAs (12 galaxies). The latter two classes are merged to form the
control sample to compare with SB galaxies. This sample, although not
statistically complete, has been selected according to the following
requirements:
\begin{list}{--}{}
   \item All objects are relatively nearby, which ensures good
   spatial resolution with a $6\arcsec$ or $3\arcsec$ pixel size (the extension
   of the central concentration is typically 5--10 pixels in
   diameter). At the distances of the sample, a $3\arcsec$ pixel corresponds
   to linear sizes between 60 and 900\,pc. Virgo galaxies, less
   extended and all imaged with $6\arcsec$ pixels in order to increase
   the signal to noise ratio, are resolved but with less detail.

   \item The sample was selected to avoid non-stellar activity as well
   as strong signs of tidal interaction, with a few exceptions in the
   {\it Cambarre} and {\it Virgo} subsamples, detailed in the following.

   \item Galaxies included in the first two programs are moderately
   inclined on the line of sight ($i \leq 50\degr$). This was not a
   requirement for the other programs, so that one third of the
   Virgo galaxies and one third of the supplementary galaxies are inclined
   by more than $60\degr$.

   \item The number of SA-SAB galaxies is comparable to that of SBs
   and both groups span the whole de Vaucouleurs spiral sequence from
   types S0/a to Sdm.

   \item Both barred and unbarred galaxies cover a
   large range of far-infrared luminosities (between $10^{8.6}$ and
   $10^{11}\:L_{\odot {\rm bol}}$), but none would be classified as
   an infrared luminous galaxy, except NGC\,7771 which is at the lower
   boundary of this class -- defined by $10^{11} < L_{\rm FIR} <
   10^{12}\:L_{\odot {\rm bol}}$ \citep{Sandersb}.
\end{list}

Another property which was not a selection criterion is that absolute blue
magnitudes are equal to or greater than the typical magnitude of the Schechter
luminosity function in the field, $M_{\rm B}^* \simeq -21$ (more exactly, they
range between -21.17 and -17.38).

Despite the incompleteness of the sample, we have checked that it is
very similar to the magnitude-limited CfA galaxy sample
\citep{Thuan}, from the point of view of its infrared brightness
normalized by blue starlight. For CfA spiral galaxies detected in all
4 IRAS bands and with blue magnitudes in the RC3, log\,($F_{\rm FIR}/F_{\rm B}$)
falls in the interval [$-0.97 ; +0.98$] with a mean value of 0.05. Using
the same IRAS references as those in \citet{Thuan}, {\it i.e.} in order of
preference \citet{Thuan}, \citet{Rice}, \citet{Soifer} and
\citet{Moshir}, galaxies in our sample have log\,($F_{\rm FIR}/F_{\rm B}$)
in the interval [$-1.58 ; +1.67$] with a mean value of 0.01\,. For that set
of references, a Wilcoxon-Mann-Whitney (WMW) test indicates that the
probability for the two populations to have the same $F_{\rm FIR}/F_{\rm B}$
distribution is about 75\%. We note that the IRAS 12\,$\mu$m fluxes
often disagree with our 7 and 15\,$\mu$m fluxes, although the bandpasses
overlap. Thus, when we use IRAS data, we take them from the references
we consider the most reliable ({\it i.e.} which provide the best match
between 12\,$\mu$m and our 7--15\,$\mu$m flux densities). In that case,
log\,($F_{\rm FIR}/F_{\rm B}$) falls in the interval [$-0.77 ; +0.95$] with a
mean value of 0.02, and the WMW test gives a probability of about 40\%.
Hence, our sample is not different from optically complete samples
regarding the fraction of the energy radiated in the infrared.

Table~\ref{tab:tab_sample} lists some general characteristics of the
galaxies. The morphological classification adopted is that of the RC3
\citep{Vaucouleurs}. Although it is based on blue images, which may not be
as appropriate as near-infrared images for detecting bars, many more galaxies
are classified as barred in this catalog than for instance in \citet{Sandage}.
We have found only two galaxies classified as SA in the RC3 and possessing
a bar (as described in the following). A drawback of using the SB and SAB
classes of the RC3 is that they do not constitute a measure of the bar
dynamical strength. The bar strength is however difficult to quantify,
and reliable measures, such as those of \citet{Buta}, are scarce.
In the following, we will refer to bar lengths, normalized by the disk
diameter, because longer bars are able to collect gas from inside a larger
area and have low axis ratios, which are among the (unsatisfactory)
quantities used to estimate bar strengths; bar lengths are in addition 
relatively easy to measure.

The two sub-samples of spirals found in the field or loose groups and
Virgo galaxies have been separated, because they differ both in their
aspect in the infrared (Virgo members are fainter and less extended)
and in their environment. Although Virgo is not a very rich cluster,
the interaction of central galaxies with the intracluster gas and with
their neighbours is likely to cause either a depletion or an enhancement
of star formation activity in the outer parts of disks and also to have
global dynamical consequences. An extreme case is the galaxy NGC\,4438
(= VCC\,1043), whose very perturbed morphological appearence was
successfully modelled by \citet{Combes} as the result of a collision
with NGC\,4435. Several Virgo members have truncated H{\scriptsize I}
disks due to the interaction with the cluster hot gas \citep{Cayattea};
a very clear example is NGC\,4569 (= VCC\,1690), which on optical
photographs shows the juxtaposition of a bright and patchy inner disk
structured by star formation sites and dust lanes, and a low surface
brightness and very smooth outer disk with faint spiral arms. Severely
H{\scriptsize I}-stripped galaxies can indeed be recognized in the optical
as anemic \citep[defined by][ as an intermediate and parallel sequence
between lenticulars and spirals]{Bergh}, due to the suppression of star
formation where the gas density is too low. Table~\ref{tab:tab_sample} also
indicates whether signatures of nuclear activity or tidal interaction exist.

In addition to these, some galaxies deserve special comments (see the
Atlas for more details) and should be considered cautiously in the
interpretation of the data set:
\begin{list}{--}{}
   \item NGC\,337, 1385 and 4027 are strongly asymmetric and fit in
   the category of magellanic barred spirals.
   \item NGC\,4691 and 1022 have an amorphous structure and highly
   centrally concentrated interstellar tracers. Their morphology is
   suggestive of merger results. Their star formation activity may
   therefore not be a consequence of the bar, since the latter was
   likely produced at the same time by the same cause, {\it i.e.} the
   merger.
\end{list}

\section{Observations and photometric results}
\label{sec:sec_obs}

All galaxies were observed with two broadband filters, LW3 (12--18\,$\mu$m)
and LW2 (5--8.5\,$\mu$m), that we shall hereafter designate by their central
wavelength, respectively 15 and 7\,$\mu$m. This was expected to provide
$F_{15}/F_7$ colors directly linked with star formation intensity, since
the LW2 filter covers the emission from a family of bands (see
Sect.~\ref{sec:sec_dust}), which are ubiquitous in the interstellar medium,
and LW3 was supposed to cover mainly a thermal continuum observed to rise
faster than the emission bands in star-forming regions, for instance from
the IRAS $F_{25}/F_{12}$ ratio \citep{Helou}; however, we will see that
the picture is more complicated. Maps covering the whole infrared-emitting
disk were constructed in raster mode. In all cases, the field of view is
large enough to obtain a reliable determination of the background level,
except for NGC\,4736 and 6744. The pixel size is either $3\arcsec$ or
$6\arcsec$, depending on the galaxy size. The half-power/half-maximum
diameters of the point spread function are respectively
$6.8\arcsec$/$\simeq 3.1\arcsec$ at 7\,$\mu$m with a
$3\arcsec$ pixel size, $9.5\arcsec$/$5.7\arcsec$ at 7\,$\mu$m with a
$6\arcsec$ pixel size, $9.6\arcsec$/$3.5\arcsec$ at 15\,$\mu$m with a
$3\arcsec$ pixel size and $14.2\arcsec$/$6.1\arcsec$ at 15\,$\mu$m with
a $6\arcsec$ pixel size. The data reduction is described in the Atlas.

Since the emission from various dust species and atomic lines is mixed in the
broadband filters (see Sect.~\ref{sec:sec_dust}), it is essential to complement
our maps with spectro-imaging data. These allow an estimate of the
relative importance of all species as a function of the location
inside a galaxy. We have thus obtained spectra between 5 and 16\,$\mu$m
of the inner disks ($3\arcmin \times 3\arcmin$ or
$1.5\arcmin \times 1.5\arcmin$) of five bright galaxies: NGC\,613, 1097,
1365, 5194 and 5236 (Fig.~\ref{fig:spectres}). Spectra
averaged over a few central pixels covering approximately the extent
of the circumnuclear region (left column) are compared with spectra
averaged over the inner disk, excluding the central part and a
possible ghost image (middle column). The right column shows the
observed spectrum of the faintest pixels, consisting of the zodiacal
spectrum contaminated by emission features from
the target galaxy, because the field of view never extends beyond the
galactic disk. For this reason, we cannot measure exactly the level
of the zodiacal foreground to remove. Instead, as explained in the
Atlas, we first fit a reference zodiacal spectrum to the average spectrum
of the faintest pixels (excluding the spectral regions where emission
features appear). The upper limit to the zodiacal foreground
is set by offsetting the fitted spectrum within the dispersion range,
with the additional constraint that the corrected disk spectrum remains
positive; the lower limit is symmetric to the upper limit with respect to
the fit. This makes little difference for the nuclear spectra but it does
for the disk spectra, although it does not affect the spectral shape.
Note that due to the configuration of the instrument, two different filters
are used for the short and long wavelength parts of the spectra, and that
a small offset can result at the junction of these filters, around
9.2\,$\mu$m.

The mid-infrared maps generally show an intense circumnuclear source.
Decomposing surface brightness profiles into a central condensation
and a disk (see details in the Atlas), we define a radius for this
circumnuclear region, $R_{\rm CNR}$. Total fluxes and fluxes inside $R_{\rm CNR}$
are listed in Table~\ref{tab:tab_photom} with the background level for each
broadband filter. Explanations about the method employed for photometry
and the estimation and meaning of errors can be found in the Atlas. The
dominant uncertainty arises from memory effects for relatively bright galaxies,
and from other sources of error (essentially the readout and photon noise)
for faint galaxies, especially at 15\,$\mu$m. For galaxies drawn from the
{\it Sf-glx} project, the number of exposures per sky position is very small
($\simeq 10$) and does not allow a proper estimate of memory effects: their
photometric errors are thus especially ill-determined. Typical errors are
$\approx 10$\% at 7\,$\mu$m and 18\% at 15\,$\mu$m. Note that flux density
calibration uncertainties, which are of the order of 5 to 10\%, are not included.
However, this is a systematic effect, hence not affecting relative fluxes.

\section{Nature of the mid-infrared emitting species}
\label{sec:sec_dust}

\begin{figure*}[!ht]
\vspace*{-0.5cm}
\begin{minipage}[t]{6cm}
\centerline{\resizebox{7cm}{!}{\rotatebox{90}{\includegraphics{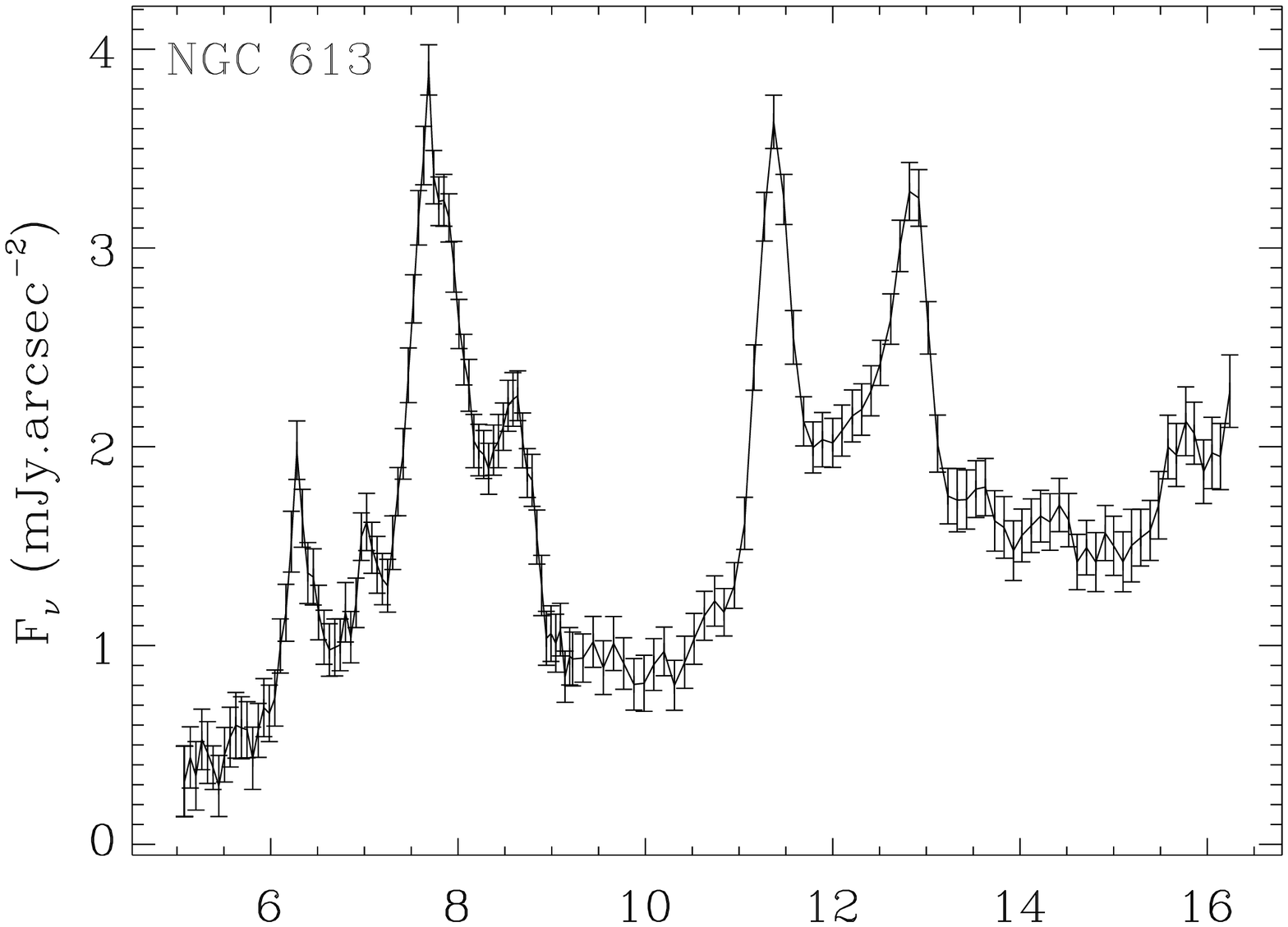}}}}
\end{minipage}
\begin{minipage}[t]{6cm}
\centerline{\resizebox{7cm}{!}{\rotatebox{90}{\includegraphics{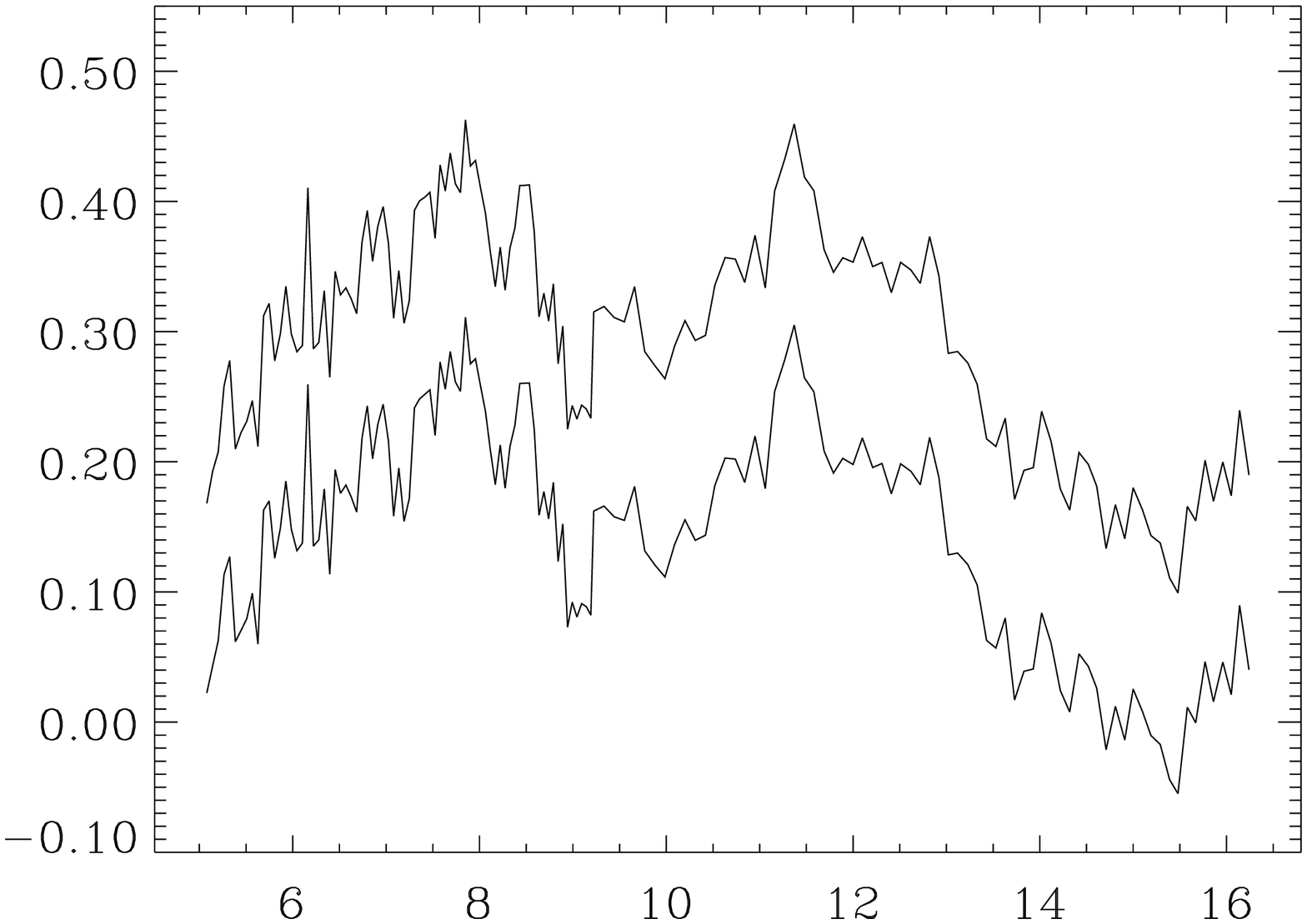}}}}
\end{minipage}
\begin{minipage}[t]{6cm}
\centerline{\resizebox{7cm}{!}{\rotatebox{90}{\includegraphics{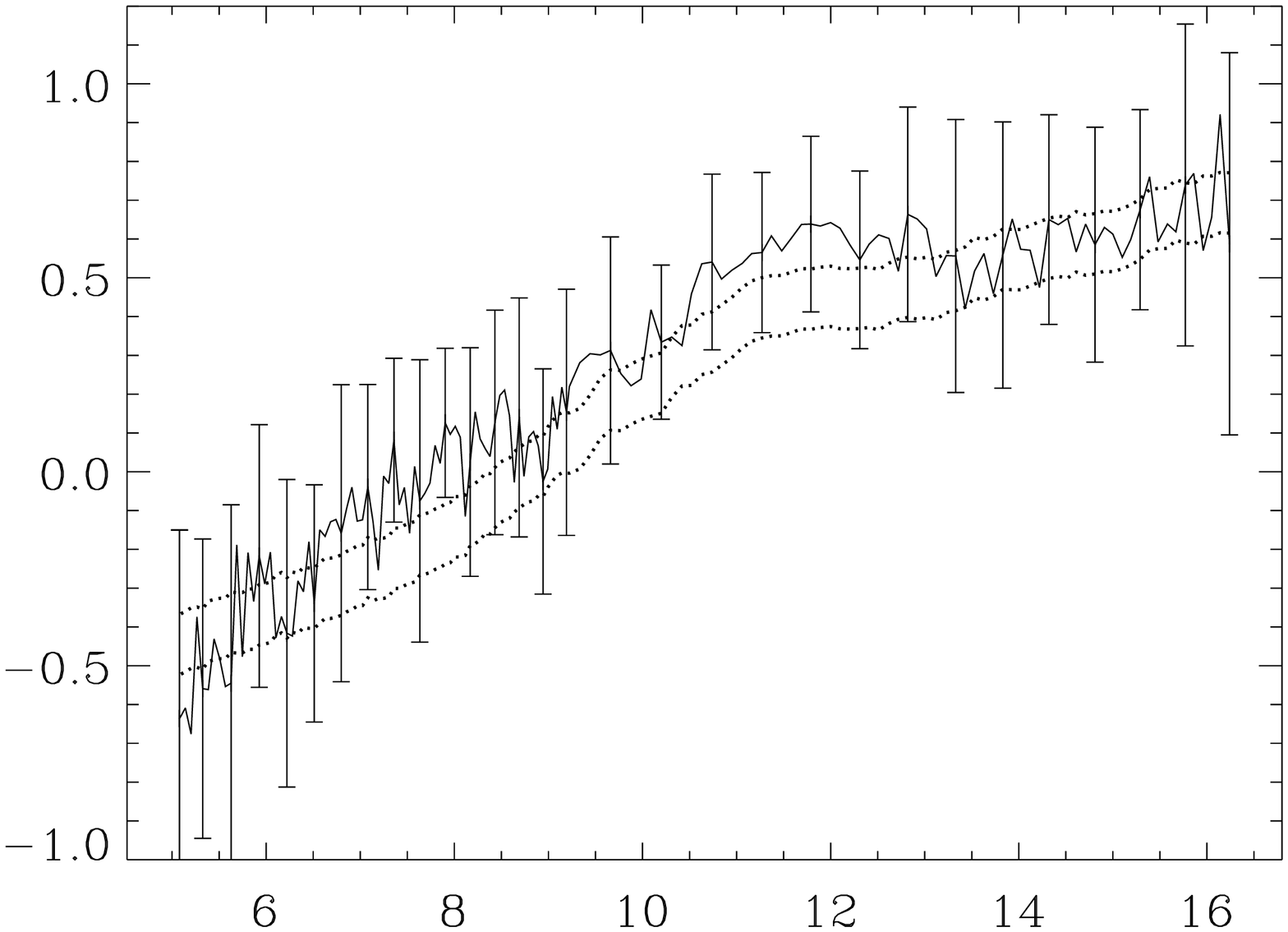}}}}
\end{minipage}
\begin{minipage}[t]{6cm}
\vspace*{-1.15cm}
\centerline{\resizebox{7cm}{!}{\rotatebox{90}{\includegraphics{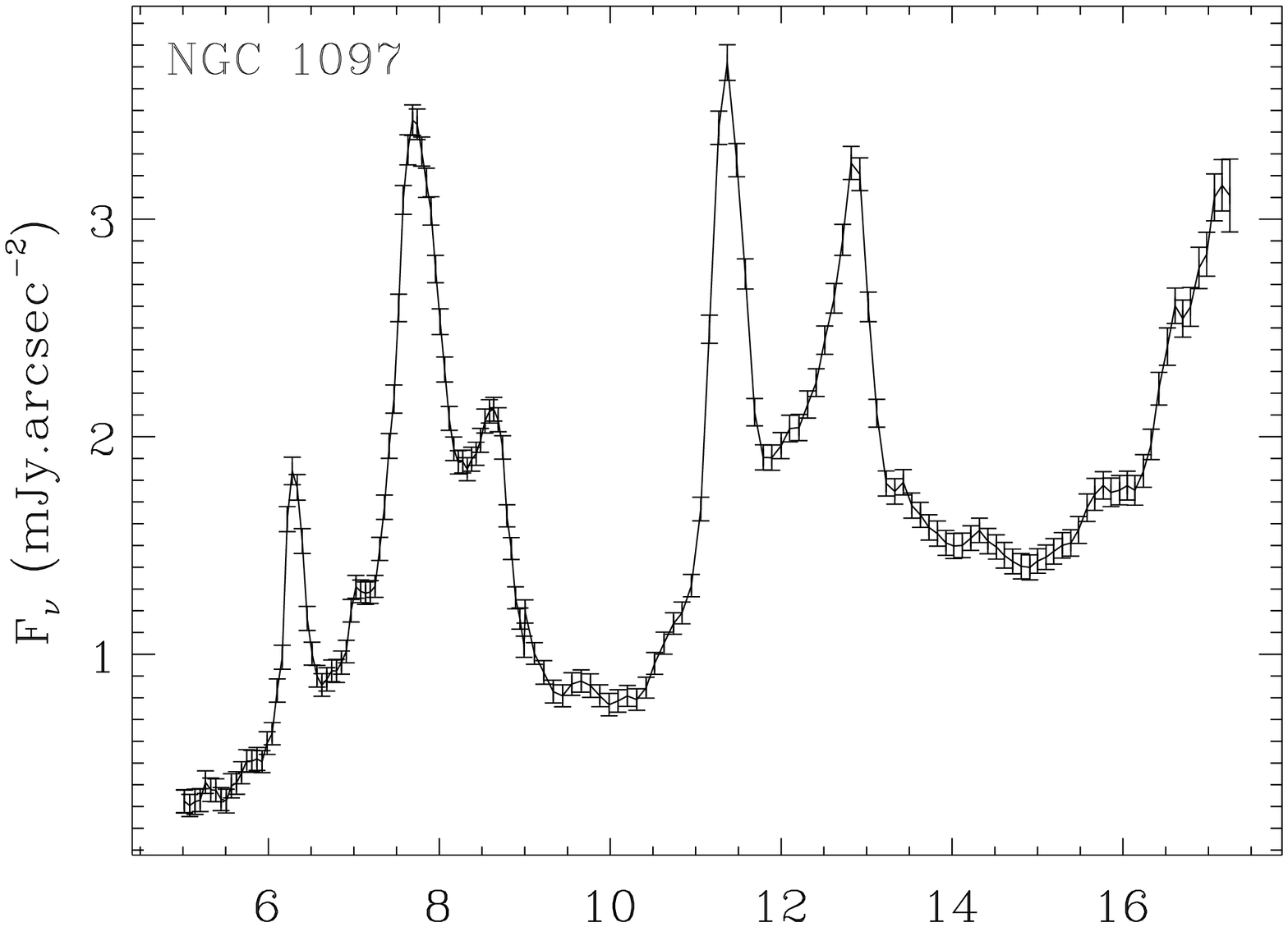}}}}
\end{minipage}
\begin{minipage}[t]{6cm}
\vspace*{-1.15cm}
\centerline{\resizebox{7cm}{!}{\rotatebox{90}{\includegraphics{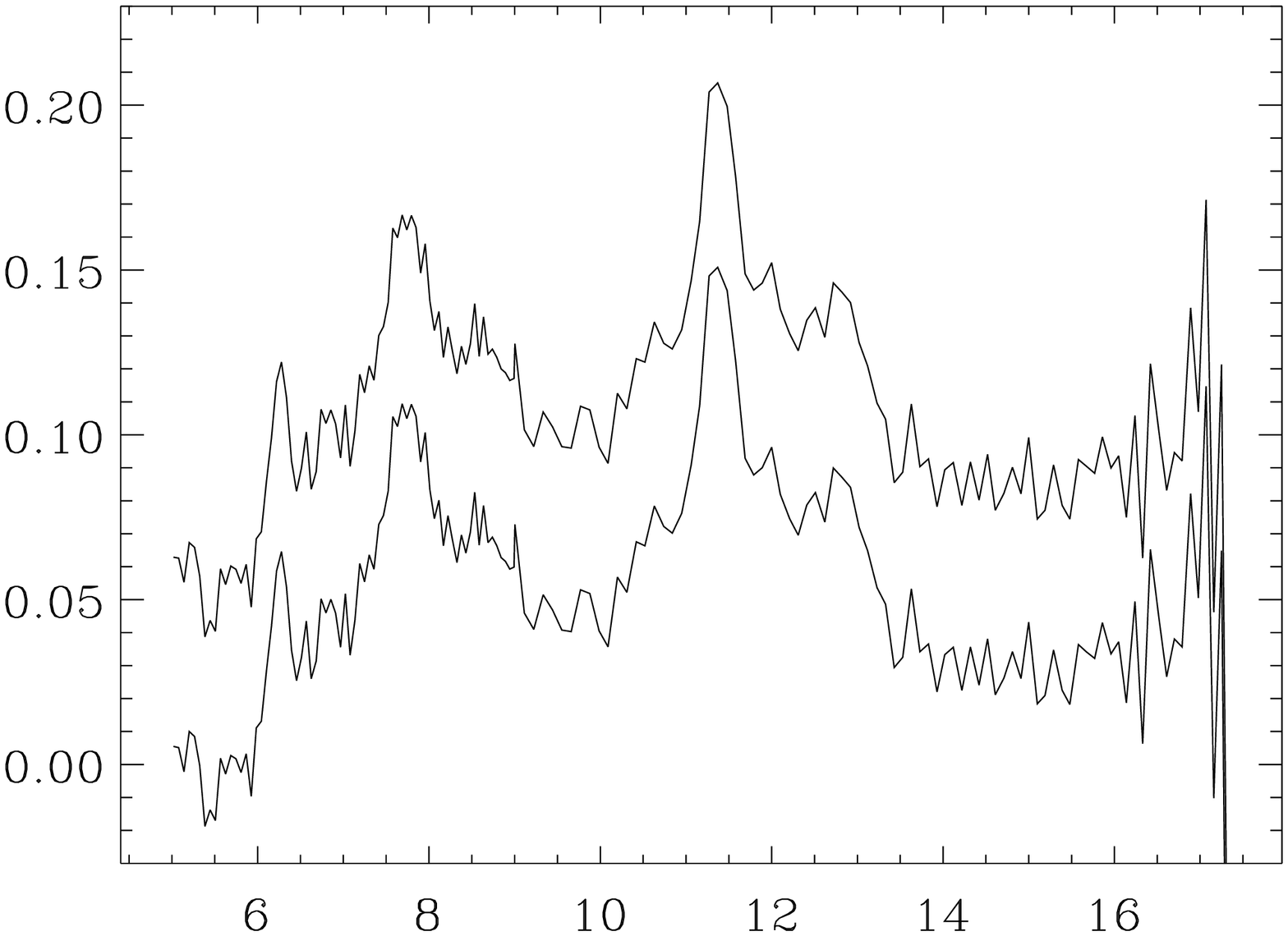}}}}
\end{minipage}
\begin{minipage}[t]{6cm}
\vspace*{-1.15cm}
\centerline{\resizebox{7cm}{!}{\rotatebox{90}{\includegraphics{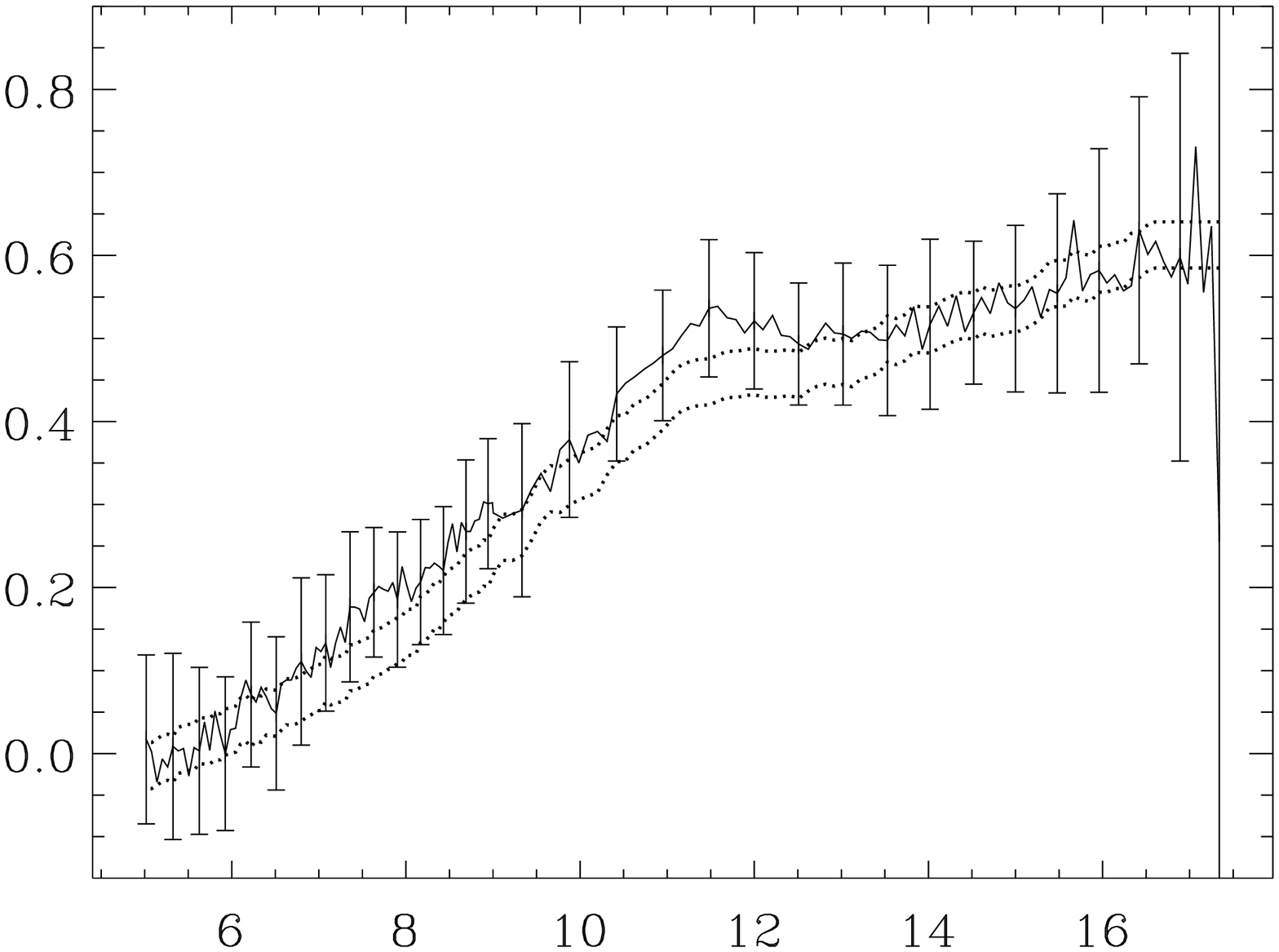}}}}
\end{minipage}
\begin{minipage}[t]{6cm}
\vspace*{-0.75cm}
\centerline{\resizebox{7cm}{!}{\rotatebox{90}{\includegraphics{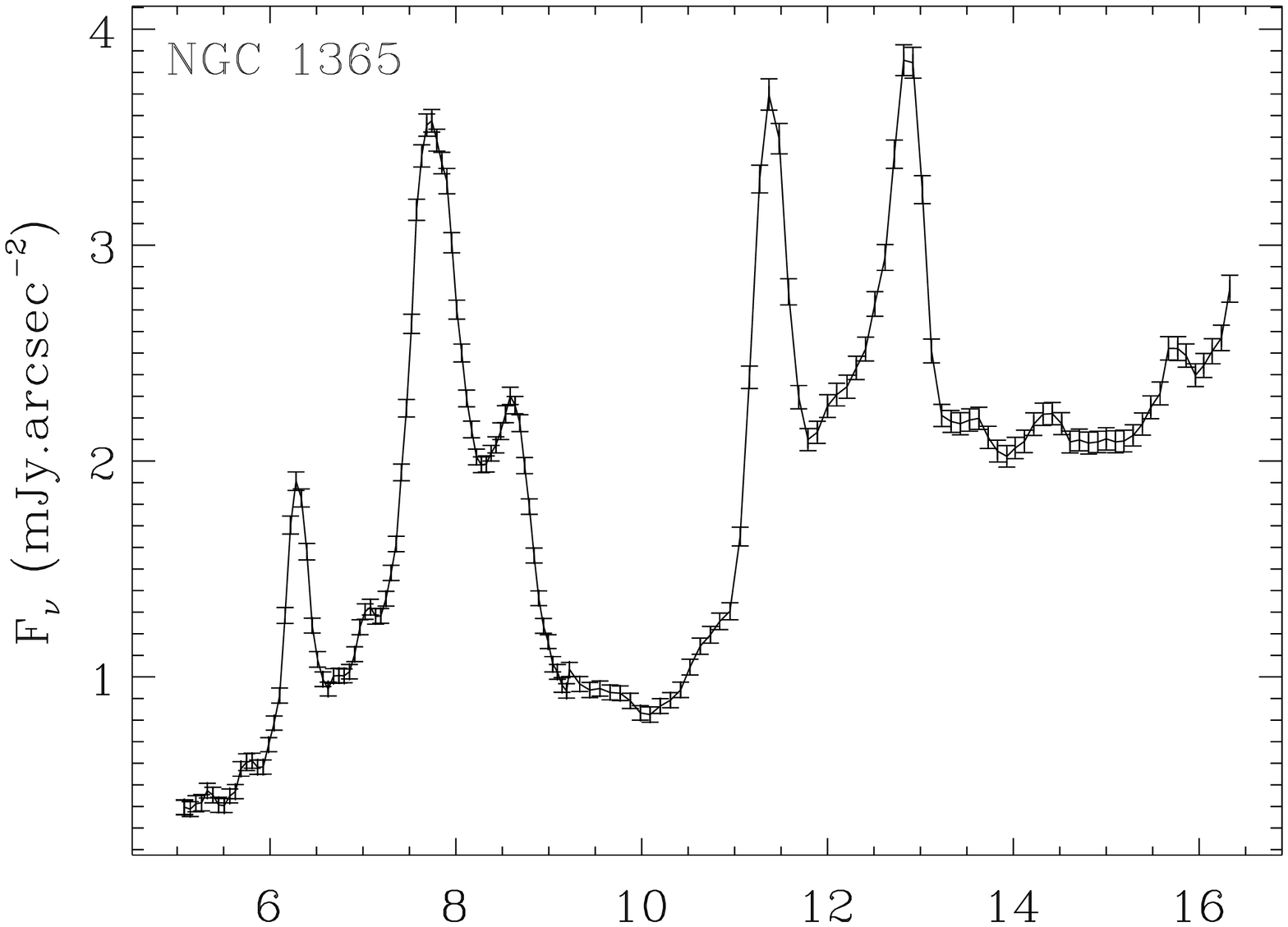}}}}
\end{minipage}
\begin{minipage}[t]{6cm}
\vspace*{-0.75cm}
\centerline{\resizebox{7cm}{!}{\rotatebox{90}{\includegraphics{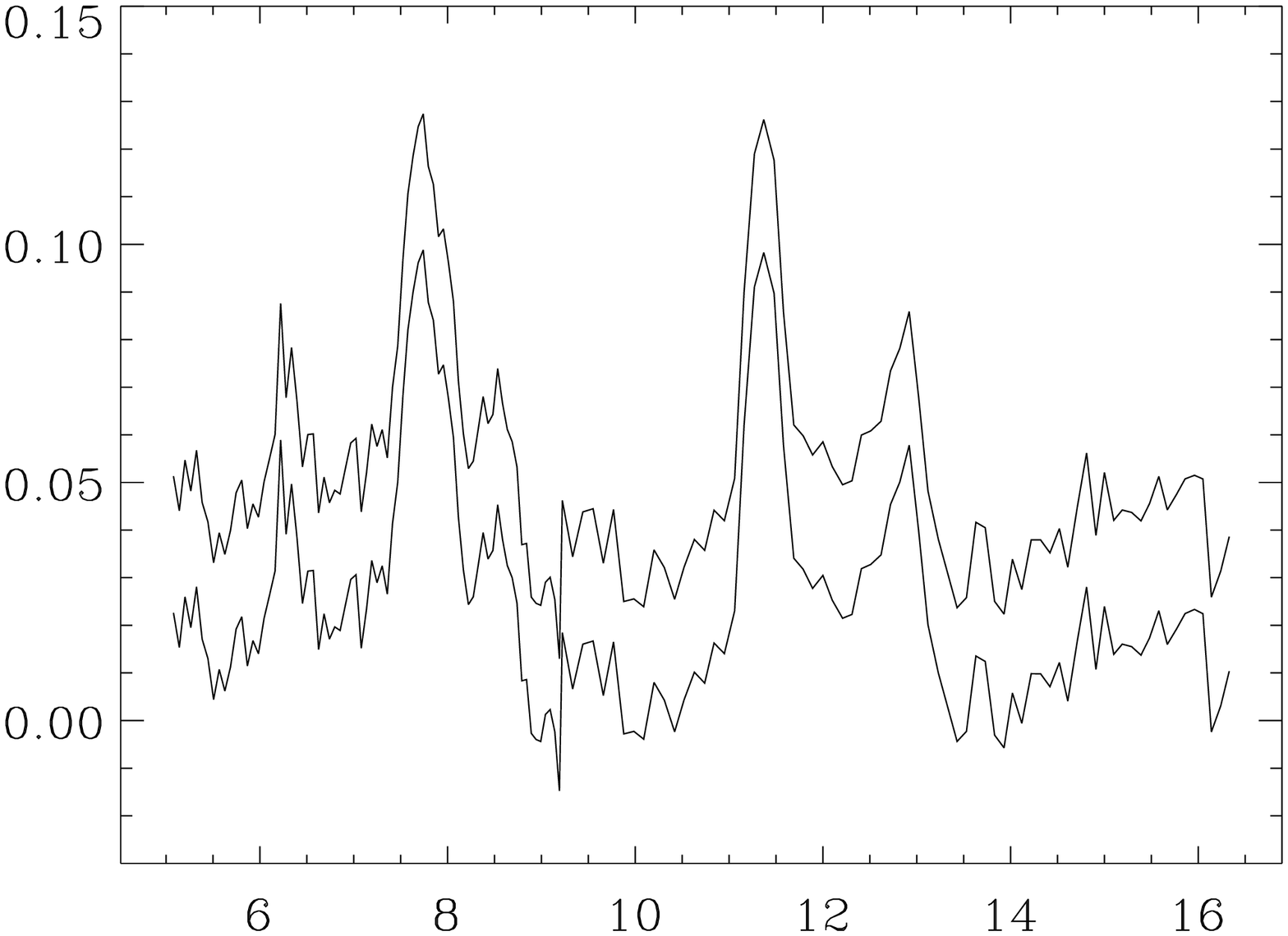}}}}
\end{minipage}
\begin{minipage}[t]{6cm}
\vspace*{-0.75cm}
\centerline{\resizebox{7cm}{!}{\rotatebox{90}{\includegraphics{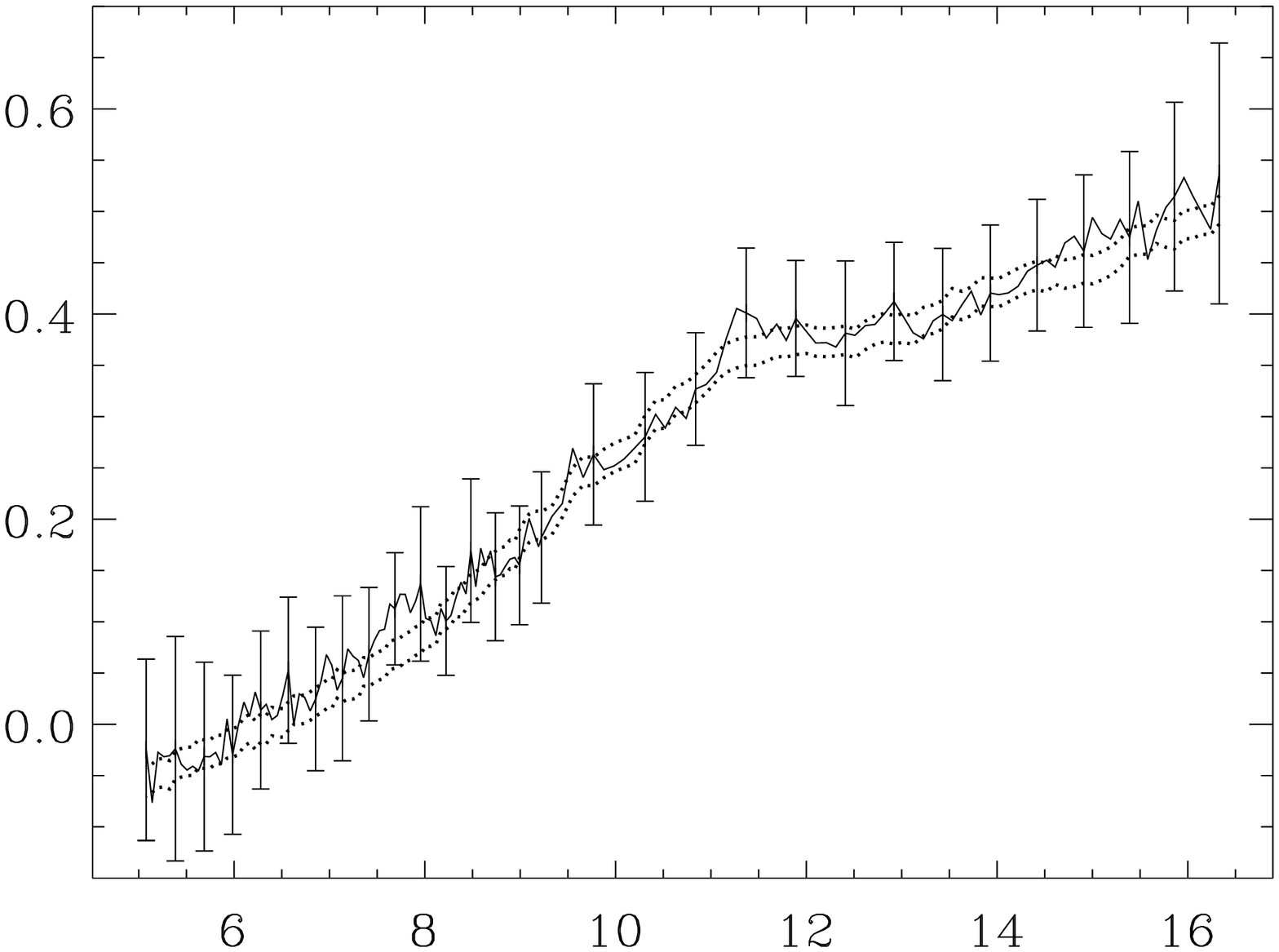}}}}
\end{minipage}
\begin{minipage}[t]{6cm}
\vspace*{-0.75cm}
\centerline{\resizebox{7cm}{!}{\rotatebox{90}{\includegraphics{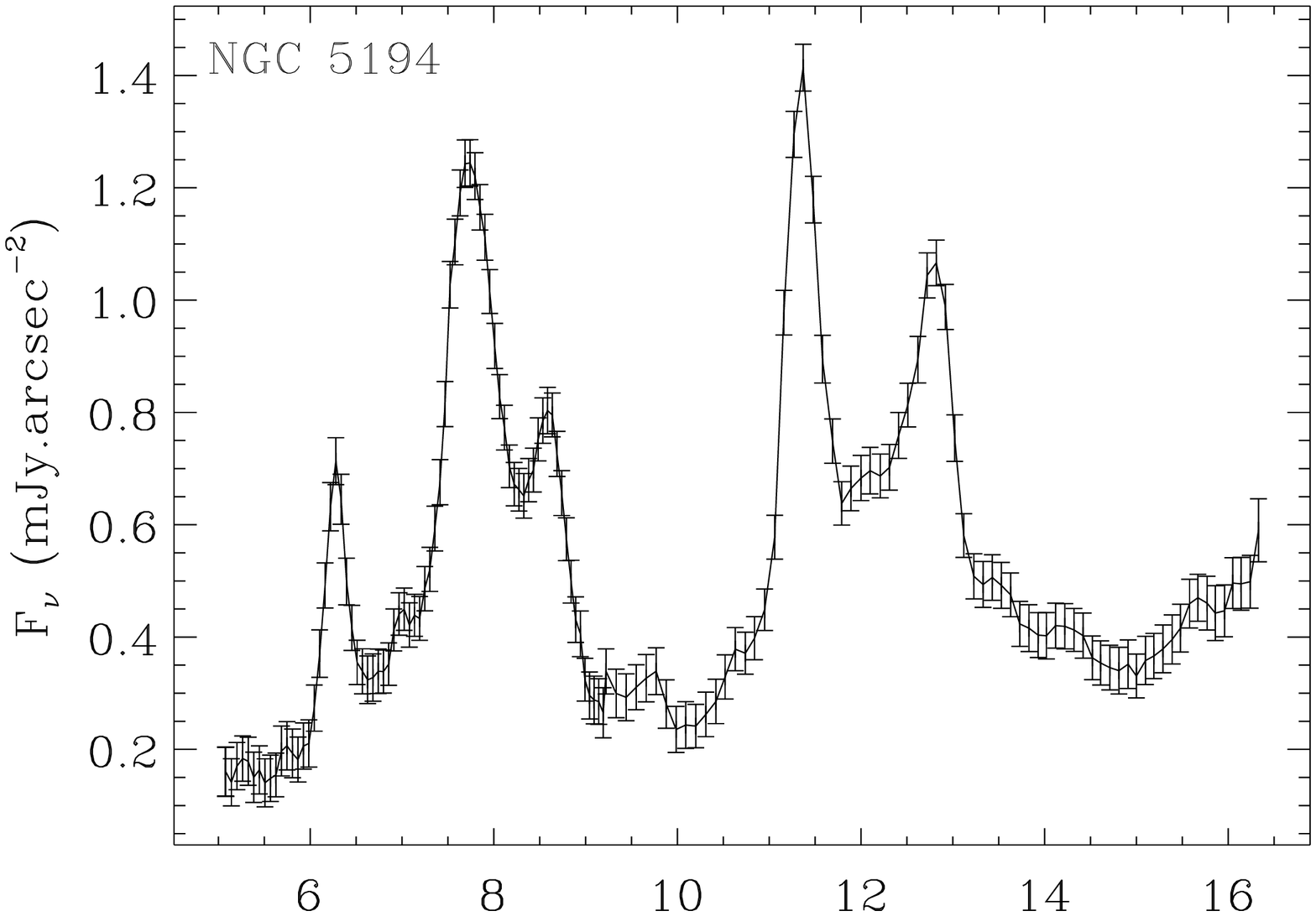}}}}
\end{minipage}
\begin{minipage}[t]{6cm}
\vspace*{-0.75cm}
\centerline{\resizebox{7cm}{!}{\rotatebox{90}{\includegraphics{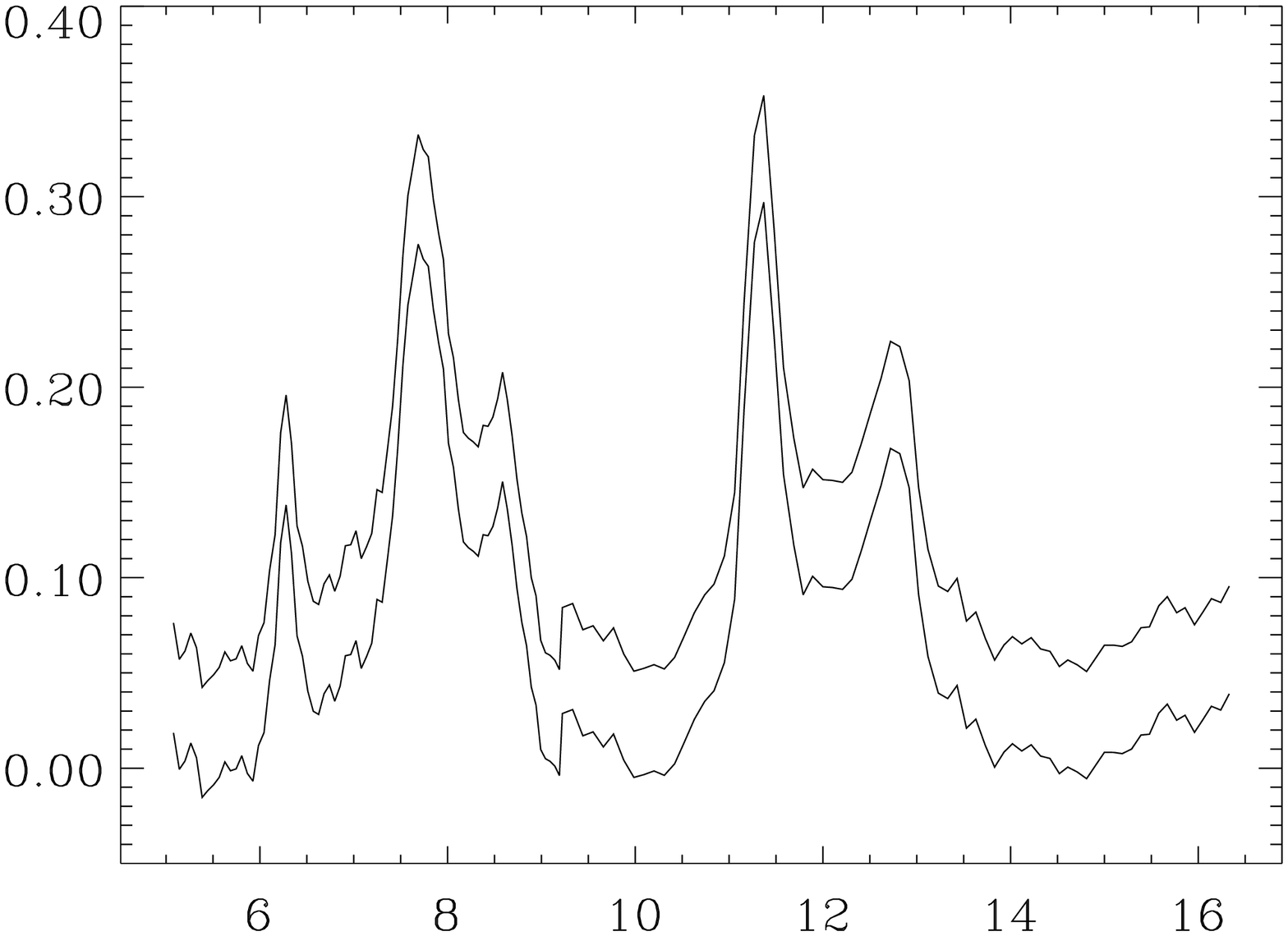}}}}
\end{minipage}
\begin{minipage}[t]{6cm}
\vspace*{-0.75cm}
\centerline{\resizebox{7cm}{!}{\rotatebox{90}{\includegraphics{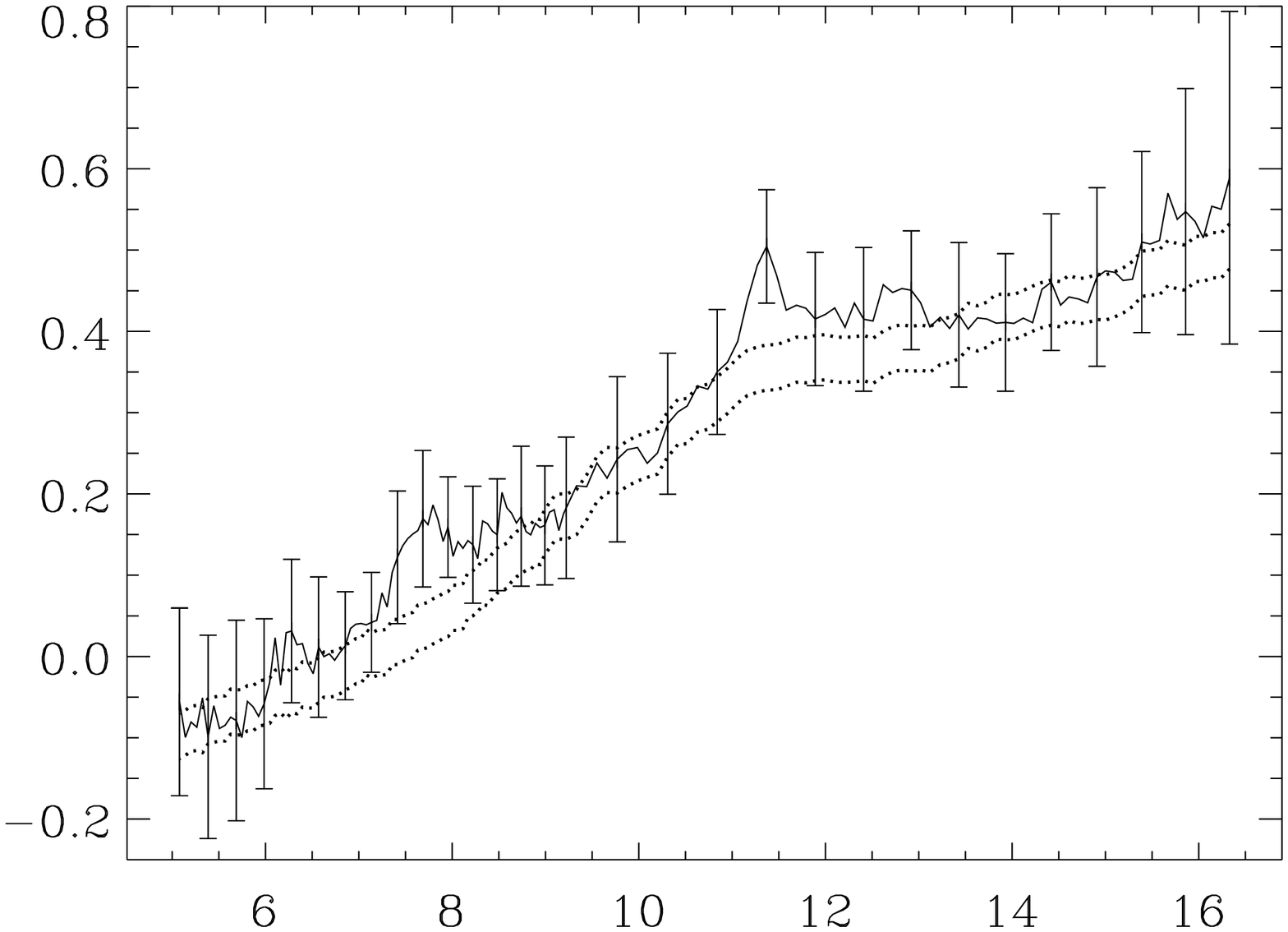}}}}
\end{minipage}
\begin{minipage}[t]{6cm}
\vspace*{-0.75cm}
\centerline{\resizebox{7cm}{!}{\rotatebox{90}{\includegraphics{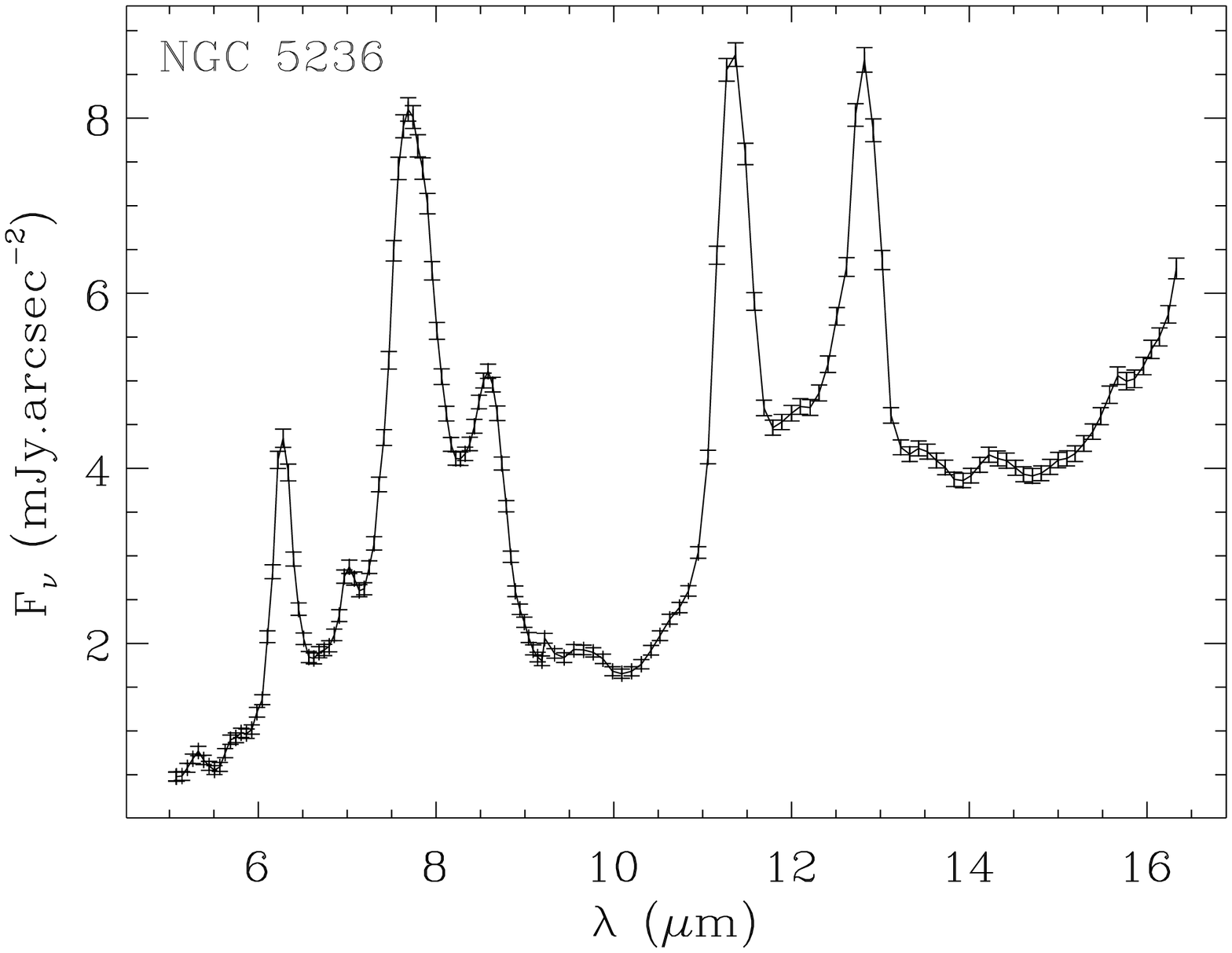}}}}
\end{minipage}
\begin{minipage}[t]{6cm}
\vspace*{-0.75cm}
\centerline{\resizebox{7cm}{!}{\rotatebox{90}{\includegraphics{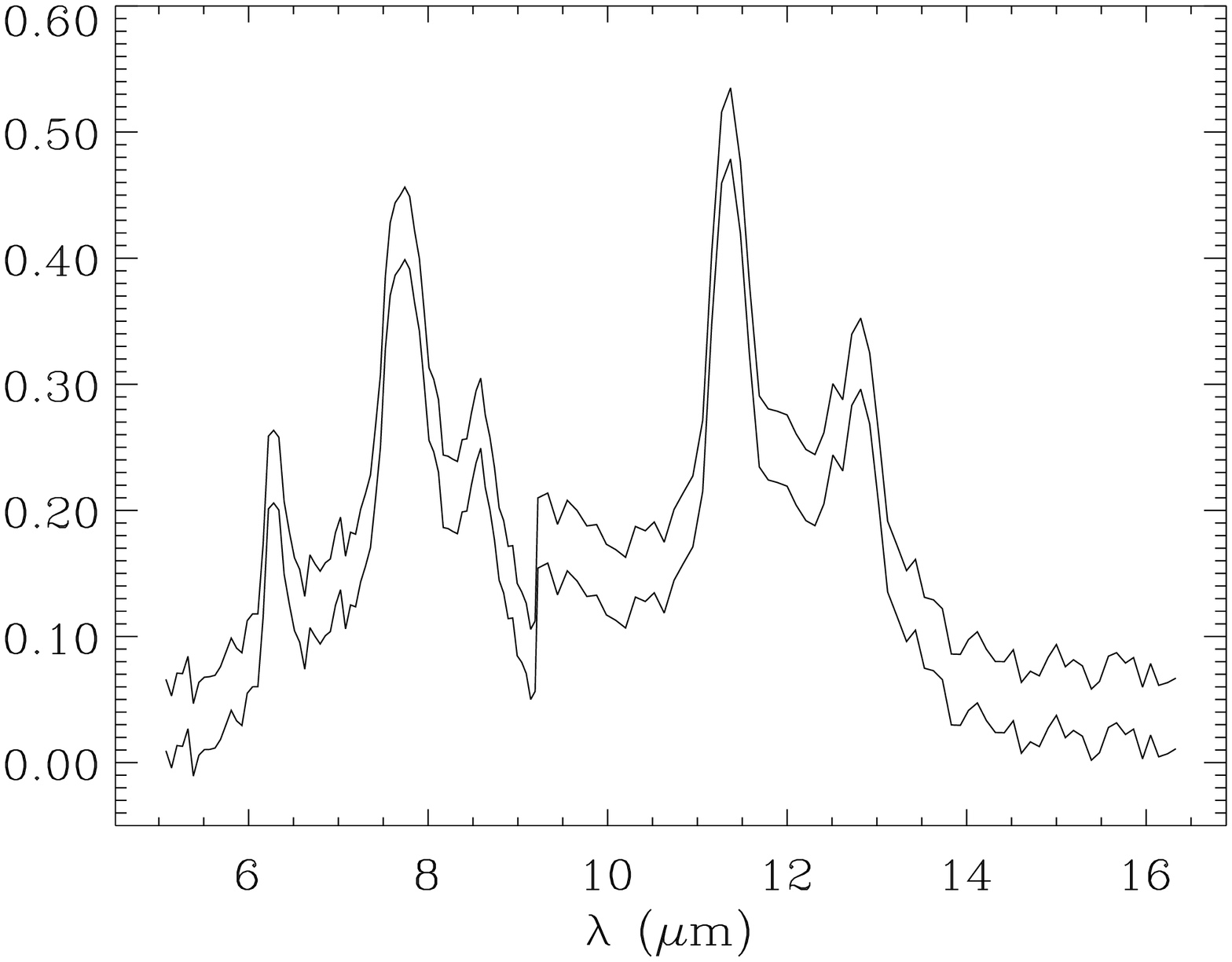}}}}
\end{minipage}
\begin{minipage}[t]{6cm}
\vspace*{-0.75cm}
\centerline{\resizebox{7cm}{!}{\rotatebox{90}{\includegraphics{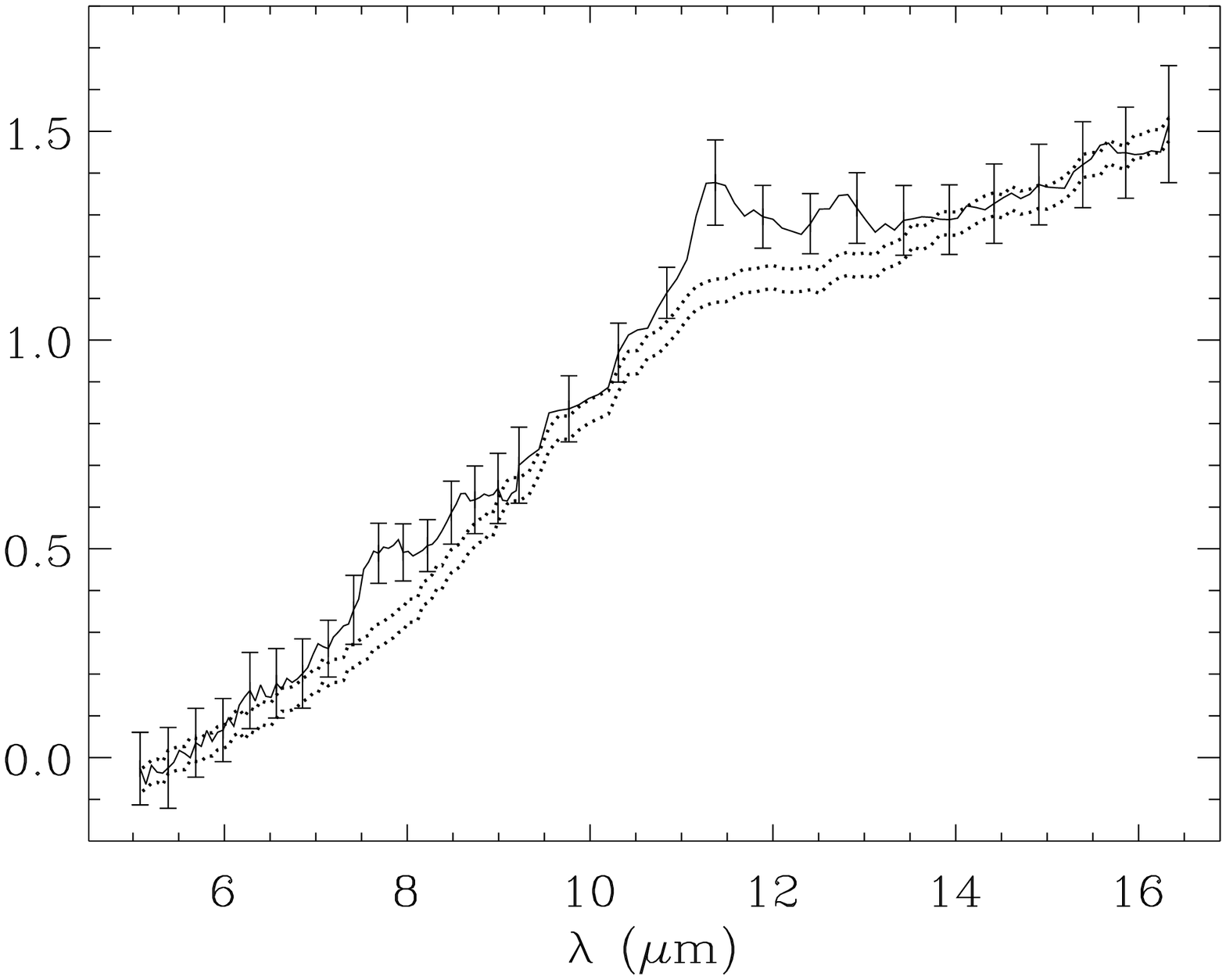}}}}
\end{minipage}
\vspace*{-0.3cm}
\caption{Spectra of central regions (left) and the inner disk (middle).
         The upper and lower limits are determined from limits on the zodiacal
         spectrum shown with dotted lines (right), adjusted using the average
         spectrum of the faintest pixels, also shown with its dispersion.
         The flux unit for all spectra is mJy\,arcsec$^{-2}$.}
\label{fig:spectres}
\end{figure*}

The spectra shown in Fig.~\ref{fig:spectres} are strikingly similar to
one another. They contain some features also seen in spectra of reflection
nebulae, atomic and molecular envelopes of H{\scriptsize II} regions, atmospheres
of C-rich evolved stars as well as the diffuse interstellar medium. We can
thus safely assume that the results obtained on these resolved
Galactic objects can be readily extrapolated to the emission of
galaxies where individual sources are no longer resolved.

The emission between 5 and 16\,$\mu$m is dominated by the so-called
unidentified infrared bands (UIBs) at 6.2, 7.7, 8.6, 11.3 and
12.7\,$\mu$m. Our spectra also display weak features which have
previously been detected as broad features in SWS spectra of starburst
objects \citep{Sturm} at {\it e.g.} 5.3, 5.7, 10.7, 12.0, 13.6, 14.3
and 15.7\,$\mu$m\footnote{We also detect a weak and unknown emission
feature between 9.3 and 9.9\,$\mu$m, which seems brighter, relatively to
UIBs, in disks than in central regions. However, the very poor signal
to noise ratio of disk spectra does not allow us to be conclusive. It
cannot be an artefact due to the change of filter since that change occurs
after the feature is observed. It is too narrow to be emitted by silicates.
The identification with ionized PAHs \citep[see {\it e.g.} ][]{Allamandola}
would be inconsistent with the fact that the flux
ratio of this feature to classical UIBs seems higher in regions of low
radiation density and excitation than in central regions. It is also
unlikely that it corresponds to the H$_2$ rotational line at
9.66\,$\mu$m since this is characteristic of warm and excited molecular
clouds in starburst nuclei \citep[{\it e.g.} ][]{Spoon}. Finally, we mention
that it also matches in wavelength a feature from the CH$_3$
functional group at 9.6\,$\mu$m \citep{Duley}.}. A 7.0\,$\mu$m
feature can tentatively be identified as an [Ar{\scriptsize II}] line
(6.99\,$\mu$m) or an H$_2$ rotational line (6.91\,$\mu$m), but our spectral
resolution ($\Delta \lambda / \lambda \approx 40$) prevents a more definite
identification. We note however that the [Ar{\scriptsize II}] line has been
identified in the high-resolution SWS spectra of starburst galaxies
\citep{Sturm}.

It was originally proposed by \citet{Duley} that UIBs are due to
organic functional groups on carbonaceous grains. \citet{Leger} instead
favoured vibration modes of C-C and C-H bonds only, in large polycyclic
aromatic molecules not in thermal equilibrium with the local radiation field
(the so-called PAH model). The constancy of the spectral energy distribution
of UIBs, regardless of the radiation field \citep{Sellgren, Uchida},
implies an impulsive heating mechanism, where upon absorption of a single
UV photon, the carriers undergo a very rapid and large temperature increase
and then radiatively cool before the next absorption. Alternative candidates
for the UIB carriers are various hydrogenated and oxygenated carbon
grains, amorphous but partially ordered at the smallest scale \citep{Borghesi,
Sakata, Papoular}, much similar to the idea of \citet{Duley}. Recent work
by \citet{Boulanger2} indicates that UIBs are not due to molecules
such as PAHs, but more likely to aggregates of several hundred atoms.

In the interstellar medium surrounding the OB association Trapezium
\citep{Roche}, the Orion bar \citep{Giard} and M17 \citep{Cesarskya, Tran},
these features are detected in the H{\scriptsize II} region and the
molecular cloud front (provided projection effects are minor), but the
emission peaks at the photodissociation interface \citep[see also][]{Brooks}.
UIB carriers are likely destroyed in H{\scriptsize II} region cores, although the
estimation of the critical radiation field necessary to obtain a significant
reduction in UIB carrier abundance still remains to be done
\citep[compare {\it e.g.} ][]{Boulanger, Boulanger3, Contursi}.

While the 7\,$\mu$m flux in spiral galaxies essentially consists of the UIB
emission, the 15\,$\mu$m filter covers the emission from mainly two dust
species: the hot tail of a continuum attributed to very small grains (VSGs)
of the order of 0.5--10~nm in size and most often impulsively heated like
UIB carriers \citep{Desert}, and also UIBs. The red wing of the
11.3\,$\mu$m band contributes little, but the band at 12.7\,$\mu$m and the
emission plateau that connects it to the 11.3\,$\mu$m band can be important;
the smaller UIB features listed above also contribute, although to a lesser
extent. When spatial resolution is high enough, the emission from
VSGs and UIB carriers can be clearly separated: around M\,17 and in the
reflection nebula NGC\,7023, the VSG continuum strongly peaks in a
layer closer to the excitation sources than the UIBs, inside the ionized
region for M\,17 \citep[][, 1996b]{Cesarskya}. Therefore, the $F_{15}/F_7$
flux ratio decreases with increasing distance from the exciting stars
of an H{\scriptsize II} region.

In the spectra of all five galaxies (Fig.~\ref{fig:spectres}), the
intensity ratios of UIBs are remarkably stable, which is a common property
of a variety of astronomical sources \citep{Cohen, Uchida}. The only
highly varying feature is the VSG continuum that is best seen longward
of 13\,$\mu$m. It has various amplitudes and spectral slopes in galactic
nuclei. It remains very modest compared to that in starburst galaxies
\citep{Tran, Sturm}, and is hardly present in averaged disks.
In Paper II, we show that the integrated mid-infrared luminosity
of normal spiral disks is dominated by the contribution from
photodissociation regions (where the UIB emission is maximum). From a
comparison with H$\alpha$ luminosities, we show that this predominance of
the photodissociation region emission results in the fact that,
when integrated over the disk, the UIB emission is a good tracer
of massive young stars.

Finally, as alluded to earlier, a number of fine-structure lines can be
present in the mid-infrared spectral range, although their contribution
to the broadband flux is always negligible in spirals. In normal galaxies,
the most prominent is the [Ne{\scriptsize II}] line at 12.81\,$\mu$m, which at
the spectral resolution of ISOCAM is blended with the UIB at 12.7\,$\mu$m.
No lines from high excitation ions such as [Ne{\scriptsize III}] at
15.56\,$\mu$m are convincingly detected, and the [Ne{\scriptsize II}] line
at 12.81\,$\mu$m is weak, since the intensity of the blend with the UIB at
12.7\,$\mu$m, relative to the isolated UIB at 11.3\,$\mu$m, is rather stable
in different excitation conditions. Some variation however exists.
To compare the strength of the [Ne{\scriptsize II}] line in our
galaxies to that observed by \citet{Forster} in the starburst galaxies
M\,82, NGC\,253 and NGC\,1808, we have measured in a similar way the flux
of the blend $F_{12.75}$ above the pseudo-continuum drawn as a
straight line between 12.31 and 13.23\,$\mu$m, and the flux of the
11.3\,$\mu$m UIB $F_{11.3}$ with its respective continuum level defined
in the same way between 10.84 and 11.79\,$\mu$m. We find that the energy
ratio $F_{12.75}/F_{11.3}$ of circumnuclear regions decreases from
0.67 in NGC\,1365 to 0.60 in NGC\,613 and 5236, 0.52 in

\clearpage

\clearpage

\noindent
NGC\,1097 and
0.47 in NGC\,5194; in the averaged inner disks of NGC\,1365, 5236 and
5194, where it is still measurable, it takes the approximate values
0.5, 0.45 and 0.4\,. These figures are much lower than those observed
in cores of starburst galaxies by \citet{Forster}, where it can reach
1.7, and argue for a generally small contribution of [Ne{\scriptsize II}] to
the spectra. Adopting as the intrinsic $F_{12.7}/F_{11.3}$ UIB energy
ratio the minimum value of $F_{12.75}/F_{11.3}$ that we measure in our
spectra, {\it i.e.} 0.4, we obtain a maximum [Ne{\scriptsize II}] equivalent
width of 0.22\,$\mu$m in the nucleus of NGC\,5236. As for the UIBs, their
equivalent widths in disks and central regions range respectively
between EW(12.7)$ = 0.3$--0.6\,$\mu$m and EW(11.3)$ = 1.2$--1.9\,$\mu$m
(these numbers do not take into account broad UIB wings that occur if
the bands are described by Lorentzians). Our estimates give circumnuclear
values for $F_{12.7}/F_{\rm [Ne{\scriptsize II}]}$ between 1.5 and 5.7,
that we can compare with the results of \citet{Sturm} in the starburst
galaxies M\,82 and NGC\,253 from their ISOSWS spectra with a high spectral
resolution ($\lambda/\Delta \lambda \approx 1500$, versus $\approx 40$
for ISOCAM). They obtain values of 0.96 and 1.32\,. The contribution
from the [Ne{\scriptsize II}] line to our spectra is thus confirmed to be
negligible with respect to starburst galaxies.

\section{The dependence of $F_{15}/F_7$ on spiral and bar types and comparison
         with IRAS results}
\label{sec:lwglobal}

In their infrared analysis, \citet{Huang} pointed out that bars are
able to significantly enhance the total star formation only in
early-type galaxies (mixing all types between S0/a and Sbc). It is
also known that bars do not share the same properties all through the
Hubble sequence: among early types, or more exactly in spirals with
large bulges, since the relationship between Hubble type and bulge to
disk ratio is far from direct \citep[{\it e.g.} ][]{Sandage, Seigar},
they tend to be longer \citep{Athanassoulaa, Martinb},
and their amplitude, with respect to that of the
underlying axisymmetric potential, tends to be higher. For instance,
\citet{Seigar}, using K band photometry to trace the stellar mass,
find that galaxies with the strongest bars have bulge to disk mass
ratios between 0.3 and 0.5\,. For larger bulges, their number of
galaxies is too low to derive any meaningful bar strength
distribution. Early-type bars host little star formation, except
near their ends and at their center, whereas late-type galaxies
generally harbor H{\scriptsize II} regions all along the bar \citep{Garciac},
which suggests that their shocks are not as strong as in early types
\citep{Tubbs}. Inner Lindblad resonances between the gas and the
density wave, which appear when there is sufficient central mass
concentration and when the bar rotates more slowly than $(\Omega -
\kappa / 2)_{\rm max}$ (where $\Omega$ is the gas circular rotation
frequency and $\kappa$ the epicyclic frequency), and which presence
induces straight and offset shocks along the bar
\citep{Athanassoulab}, are also typically expected in early-type
galaxies. These structural differences have consequences on the
efficiency of bars to drive massive inward gas flows.

\begin{figure}[!t]
\centerline{\resizebox{9.5cm}{!}{\rotatebox{90}{\includegraphics{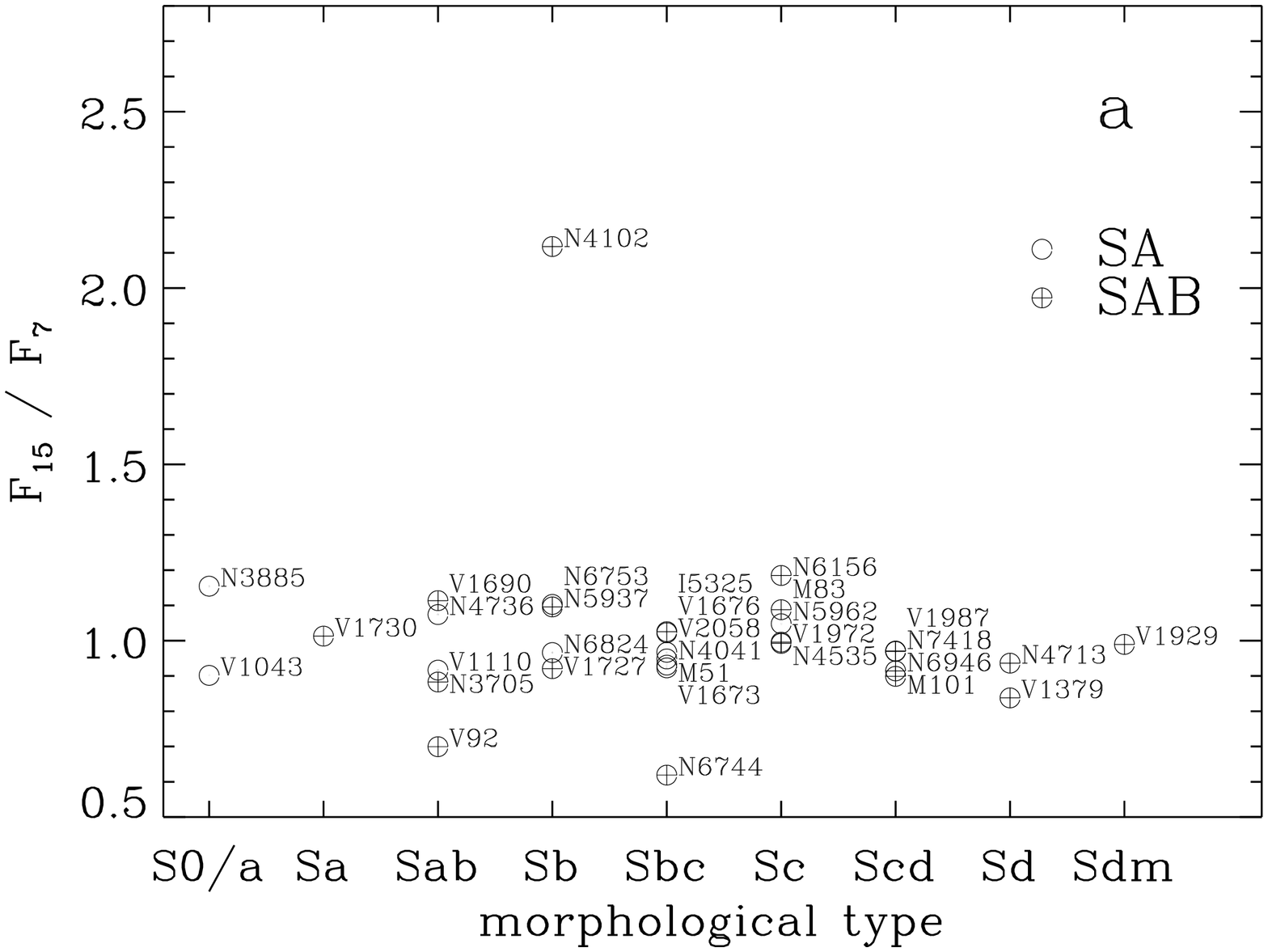}}}}
\centerline{\resizebox{9.5cm}{!}{\rotatebox{90}{\includegraphics{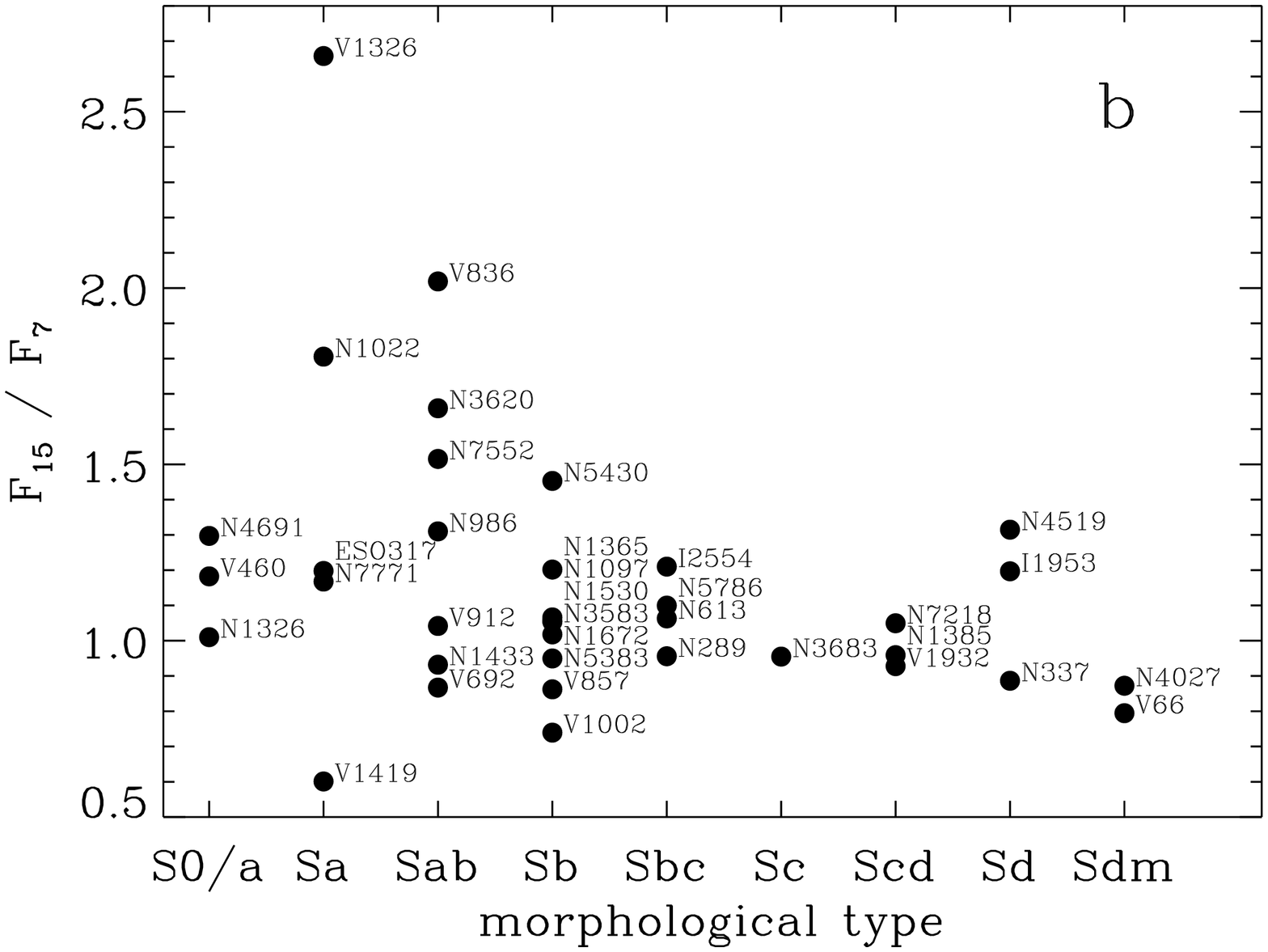}}}}
\caption{{\bf (a):} Integrated mid-infrared color $F_{15}/F_7$ as a function of
         morphological type for unbarred or weakly barred galaxies, represented
	 respectively by open circles and crossed circles. Virgo galaxies are
	 identified by their VCC number (see Table~\ref{tab:tab_sample})
         and others by their NGC number.
	 {\bf (b):} Same as {\bf (a)} for strongly barred galaxies.}
\label{fig:lwcolor_type}
\end{figure}

We show in Fig.~\ref{fig:lwcolor_type}a the distribution of $F_{15}/F_7$
according to morphological type (as given in the RC3) for the control
subsample including only SA and SAB galaxies. For this population --
excepting NGC\,4102\footnote{This galaxy shows a peculiar structure,
with a central lens-like body or fat oval of moderate length ($D_{\rm
bar} / D_{25} \approx 0.2$) surrounded by an external pseudo-ring
probably associated with a Lindblad resonance. NGC\,4102 is thus a
genuine weakly barred galaxy, and not a SB. However, such a
dynamical structure is still efficient to drive inward mass transfer.
We also point to the very strong concentration of its mid-infrared
emission (see Table~\ref{tab:tab_photom}), a property common to
early-type barred spirals (Fig.~\ref{fig:cfrac15_type}).} --, the
mid-infrared color is remarkably constant around a value of 1 (ranging
from 0.7 to 1.2). This is rather typical of the color of the surface
of molecular clouds exposed to radiation fields ranging from that
observed in the solar neighborhood to that found in the vicinity of
star-forming regions. $F_{15}/F_7$ colors observed toward H{\scriptsize II}
regions are typically of the order of 10, while those of photodissociation
regions range between 2 and the H{\scriptsize II} region values \citep{Tran}.
The fact that $F_{15}/F_7$ remains of the order of 1 in most galaxies --
it also shows generally little variation from pixel to pixel in disks
-- indicates that, at our angular resolution, emission from H{\scriptsize II}
regions and their immediate surroundings is diluted by the
larger neighboring interstellar medium (at a mean distance of 20\,Mpc,
$3\arcsec$ represent 300\,pc). In fact, in the Atlas, we show that
even in giant star-forming complexes that can be identified in the
maps, $F_{15}/F_7$ rarely exceeds 2--3.

The case of strongly barred spirals is more complex
(Fig.~\ref{fig:lwcolor_type}b): whereas many of them share the same
integrated colors as their unbarred counterparts, an important
fraction shows a color excess, the maximum color being above 2.5
instead of 1.2 for SA(B)s. Furthermore, such an excess occurs only
among the earliest morphological types, from SB0/a to SBb.
Note that in bulges, the envelopes of K-M stars can contribute an
important fraction of the mid-infrared emission. However, this would
be negligible at 15\,$\mu$m and mostly affect the 7\,$\mu$m band:
correcting for such an effect would only re-inforce the observed
trend. We also qualify that observation by noting that two galaxies,
NGC\,1022 and NGC\,4691, have
likely experienced a merger; gas may therefore have sunk to the center
as a result of the violent energy dissipation in the merger, and not
simply under the influence of the bar, which actually may have been
formed during the interaction. Dismissing these two objects however
does not change the fact that the color distribution of the strongly
barred galaxies shows 15\,$\mu$m excesses that are absent from that of
weakly barred or unbarred spirals.

One can wonder whether cluster galaxies introduce a bias in our
sample, because a number of them are perturbed by their environment
and thus may have an uncertain morphological type. \citet{Koopmann}
have shown that a significant fraction of early-type spirals in Virgo
have been ``misclassified'' due to their dearth of star formation in
the disk. The degree of resolution of spiral arms into star formation
complexes is indeed one of the three criteria defining the Hubble
sequence, but it is not unambiguously linked to the bulge to disk
ratio. Concerning several Virgo members of our sample, the bulge is very
small for the attributed type \citep{Sandage}, in such cases defined
mostly by the disk appearance. This is of course related to the anemia
phenomenon, due to gas deficiency caused by interaction with the
intracluster medium. Of our Virgo galaxies of types S0/a--Sb,
10/14 are H{\scriptsize I}-deficient, versus 3/9 for types Sbc--Sdm (see
Table~\ref{tab:tab_sample}, where $def > 1.2$ has been adopted as the
criterion for H{\scriptsize I} deficiency). This apparent segregation with
morphological type certainly results from the above classification
bias. Thus, differentiating galaxies in Fig.~\ref{fig:lwcolor_type}
according to their true bulge to disk ratio would cause an
under-representation of SA-SAB early-type spirals, which make the
crucial part of our comparison sample. If we had to discard
completely the early-type SA-SAB subsample, the maximum allowed
conclusion from Fig.\ref{fig:lwcolor_type} would be that we observe a
color excess in a fraction of early-type strongly barred galaxies,
without excluding the possibility of such an excess in early-type
non-barred galaxies, in which case another mechanism for mass transfer
would have to be thought of.

However, at least five early-type SA-SAB spirals remain which are not
H{\scriptsize I}-deficient and thus unlikely to suffer from the above bias,
namely VCC\,92 = NGC\,4192, NGC\,3705, NGC\,4736, NGC\,5937 and NGC\,6824.
We do not consider NGC\,3885, SA0/a in the RC3, because it looks
like a genuine barred galaxy: its bulge is elongated in a direction
distinct from the major axis of outer isophotes and crossed by dust
lanes; it is furthermore classified as such by \citet{Vorontsov} and
\citet{Corwin}. These galaxies show no global color excess, like the
rest of the SA-SAB subsample, and like a number of bona-fide early-type SB
galaxies with normal H{\scriptsize I} content. Hence, our view should not be
too strongly distorted by the classification bias. We have also
checked the influence of this bias on H{\scriptsize I}-deficient barred galaxies
with a color excess. On optical images, VCC\,836 = NGC\,4388
unambiguously resembles classical early-type spirals, with a prominent
bulge crossed by thick dust lanes; VCC\,460 = NGC\,4293 stands
between H{\scriptsize I}-deficient and H{\scriptsize I}-normal galaxies,
and also has an early-type aspect. The case of VCC\,1326 = NGC\,4491 is not
that clear, because it is a low-mass galaxy.

Our sample thus confirms and extends to the ISOCAM bands a phenomenon
that was evidenced from IRAS observations by \citet{Hawarden} and
\citet{Huang}, namely that a significant fraction of SB galaxies can
show an excess of 25\,$\mu$m emission (normalized to the emission at 12
or 100\,$\mu$m) compared with SA and SAB galaxies. The case of SAB
galaxies is in fact unclear: some of them show such an excess
according to \citet{Hawarden}, but they are indistinguishable from SAs
in the analysis of \citet{Huang}. From the present ISOCAM data, it
already appears that indeed SA and SAB galaxies share similar
mid-infrared properties.

\begin{figure}[!t]
\centerline{\resizebox{9.5cm}{!}{\rotatebox{90}{\includegraphics{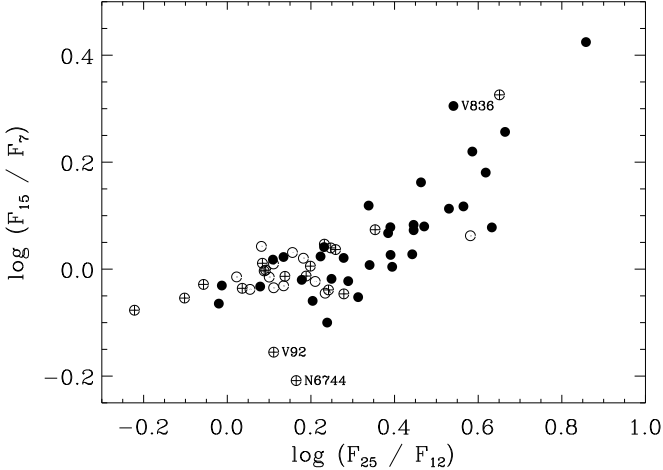}}}}
\caption{Comparison of mid-infrared colors from ISO ($F_{15}/F_7$) and IRAS
         ($F_{25}/F_{12}$). $F_{25}$ always contains the VSG emission (see
	 Sect.~\ref{sec:sec_dust}) whereas at low temperatures, $F_{15}$ is
	 dominated by UIBs, which explains the constancy of $F_{15}/F_7$ below
	 a threshold of $F_{25}/F_{12} \simeq 2$.
	 The same convention as in Fig.~\ref{fig:lwcolor_type} applies for the
         representation of SA, SAB and SB classes. We have indicated the names of
         the Sy2 galaxy NGC\,4388 = VCC\,836 (see Sect.~\ref{sec:seyfert}), and of
         the two galaxies with the lowest $F_{15}/F_7$ colors (note that NGC\,6744
         was not entirely mapped and that its integrated color is likely a
         lower limit).}
\label{fig:lwcolor_iras25s12}
\end{figure}

In order to compare more directly our results with IRAS-based results,
Fig.~\ref{fig:lwcolor_iras25s12} shows the relationship between
$F_{15}/F_7$ and $F_{25}/F_{12}$. For log\,$(F_{25}/F_{12}) < 0.3$ (which
is close to the value given by \citet{Hawarden} as the limit for the
presence of a 25\,$\mu$m excess), $F_{15}/F_7$ shows no systematic variation
and SA, SAB and SB galaxies are well mixed. Above that threshold, SB
galaxies strongly dominate (the SAB galaxy with high colors is NGC\,4102)
and the $F_{15}/F_7$ ratio follows the increase of $F_{25}/F_{12}$.
Given the nature of dust components whose emission is covered by the
7\,$\mu$m to 25\,$\mu$m filters (Sect.~\ref{sec:sec_dust}), this behavior
can be explained as follows. The classical interpretation for the
variation of $F_{25}/F_{12}$ is that it increases with the radiation
field due to the stronger contribution of VSGs to the 25\,$\mu$m than to
the 12\,$\mu$m emission, which collects mostly UIB emission
\citep{Desert, Helou}. The fact that the $F_{15}/F_7$ ratio
remains insensitive to the variation of $F_{25}/F_{12}$ for
log\,$(F_{25}/F_{12}) < 0.3$ implies that in this regime, VSGs provide
little flux to both ISOCAM bands as well. Past this threshold, the
increase of $F_{15}/F_7$ signals that the VSG continuum has entered the
15\,$\mu$m bandpass and contributes an ever increasing fraction.

The galaxies with a 15\,$\mu$m excess ($F_{15}/F_7$ above 1.2, or 0.08\,dex)
also distinguish themselves from the rest of our sample by having on
average larger far-infrared to blue luminosity ratios. For this
subsample, $L_{\rm FIR}/L_{\rm B}$ spans the range [$0.6 ; 7.3$] with a logarithmic
mean of 2.2 and dispersion by a factor 2.2, while $L_{\rm FIR}/L_{\rm B}$ of the
complementary subsample falls in the interval [$0.2 ; 9.0$], has a
logarithmic mean of 0.9 and dispersion by a factor 2.4\,. However,
the 15\,$\mu$m-excess galaxies have far-infrared luminosities that are
equivalent to those observed in the rest of the sample. Hence, in
these galaxies with a VSG emission excess, a higher fraction of the
total emission is reprocessed in the whole infrared range. There is
also a slight difference, although not statistically significant,
between SBs with no 15\,$\mu$m excess and SA-SAB galaxies: the
$L_{\rm FIR}/L_{\rm B}$ logarithmic means and dispersion factors are 1.1 and
2.7 for SBs with no excess, and 0.8 and 2.2 for SAs-SABs.

That mid-infrared color excesses occur only in SB galaxies indicates
that somehow, a global increase of the interstellar radiation field
intensity is linked to the presence of a strong bar, although this
condition is clearly not sufficient. The fact that many barred
galaxies earlier than SBb appear very similar in their integrated
color to their unbarred counterparts means that no simple link exists
between the bar class, the bulge-to-disk ratio and the onset of a
starburst in normal spirals. Several intervening parameters can be
thought of: the true strength of the bar in dynamical terms \citep[the
separation into SB and SAB classes is subjective and too rough, and in
a recent study,][ show that the SB class includes a wide
range of actual bar strengths]{Buta}; the available gas content inside
corotation; the star formation efficiency along bars and in central
regions; the timescales for starburst activation and exhaustion;
interaction with a companion or with the intracluster gas. Some of
these effects can be investigated in the present sample. We will
discuss them in Sect.~\ref{sec:orig}, but first we turn our attention
to mid-infrared properties of the central regions, as defined in
Sect.~\ref{sec:sec_obs} and in the Atlas.

\section{The role of central regions}
\label{sec:sec_rolecr}

As the presence of a bar is expected to influence the star formation
in the circumnuclear region and much less in the disk (except in the
zone swept by the bar), we are naturally led to emphasize the relative
properties of nuclei and disks. Maps shown in the Atlas demonstrate
that central regions, observed in the infrared, are prominent and
clearly distinct from other structures, much more than on optical
images.

\begin{figure}[!b]
\centerline{\resizebox{9.5cm}{!}{\rotatebox{90}{\includegraphics{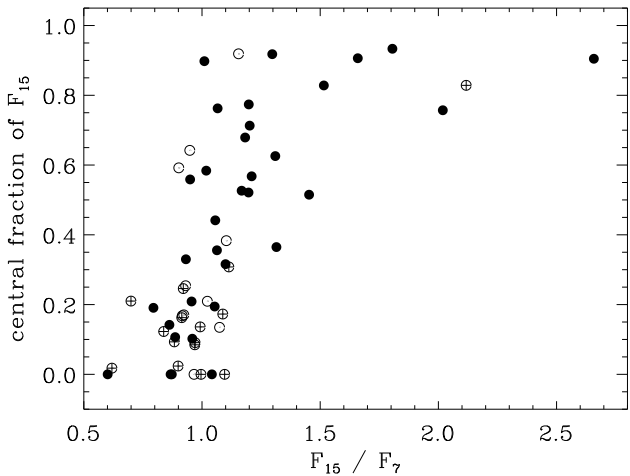}}}}
\caption{Relationship between the central flux fraction at 15\,$\mu$m and the
         integrated $F_{15}/F_7$ color. Galaxies with no identifiable central
         regions, {\it i.e.} surface brightness profiles consistent with a single
         disk component, have been placed at a null ordinate.}
\label{fig:cfrac15_color}
\end{figure}

In Fig.~\ref{fig:cfrac15_color} we plot the fraction of the total
15\,$\mu$m flux originating from the central region (inside the radius
$R_{\rm CNR}$) as a function of the global $F_{15}/F_7$ color. Galaxies for
which a central region could not be defined on the mid-infrared brightness
profiles are also shown, and are attributed a null central fraction. Galaxies
are not distributed at random in this plot, but rather on a two-arm
sequence that can be described in the following way: (1) high $F_{15}/F_7$
colors are found exclusively in systems where a high fraction of the
flux is produced in the circumnuclear regions; (2) galaxies with small
$F_{15}/F_7$ ratios ($< 1.2$) are found with all kinds of nuclear
contributions.

The bar class appears to play a part in the location of galaxies in
this diagram, although this is not clear-cut: all galaxies with high
circumnuclear contribution ($> 40$\%) and large $F_{15}/F_7$ colors
($>1.2$) are SB galaxies, apart from NGC\,4102, while SA-SAB galaxies
are quite indistinguishable from one another and cluster in the small
nuclear contribution ($< 30$\%) and low $F_{15}/F_7$ color corner of the
graph. There is also a clear preponderance of SB galaxies in all the
centrally dominated range. Only two SA-SAB galaxies show very high
concentration fractions, NGC\,3885 and NGC\,4102. The latter galaxy was
already discussed; for NGC\,3885, strong indications exist that
its bar class is incorrect (see the discussion in
Sect.~\ref{sec:lwglobal}).

However, it is quite significant that SB galaxies cover both sequences
in Fig.~\ref{fig:cfrac15_color} and in particular are found all through the
sequence of varying flux concentration and low $F_{15}/F_7$ color.
Therefore, Fig.~\ref{fig:cfrac15_color} shows that high global $F_{15}/F_7$
colors require that the flux concentration be high, and that the galaxy be
SB, but none of these two properties is enough to predict that the global
$F_{15}/F_7$ ratio will be high. To understand the importance of the flux
concentration, let us first study separately
the colors of central regions and those of disks.

\begin{figure}[!t]
\centerline{\resizebox{9.5cm}{!}{\rotatebox{90}{\includegraphics{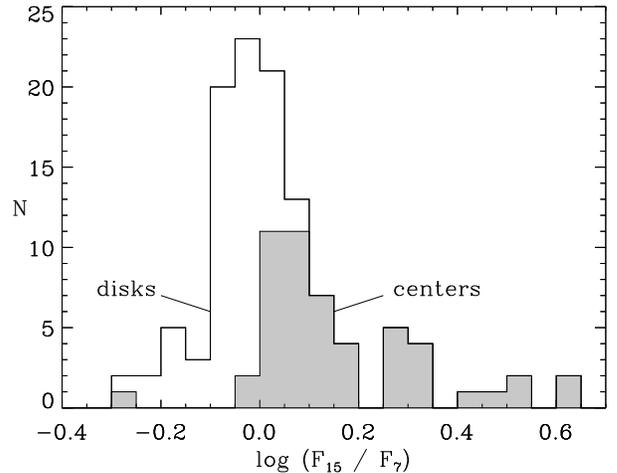}}}}
\caption{Compared histograms of $F_{15}/F_7$ colors averaged in disks and in
         circumnuclear regions. The galaxies used are respectively those whose
         disk is not strongly contaminated by the central component (the excluded
         galaxies are NGC\,1022, NGC\,4691, VCC\,1419 = NGC\,4506, NGC\,1326 and
         NGC\,3885), and those with central regions that could be adjusted on
         surface brightness profiles (otherwise the galaxy is considered to
         be composed only of a disk). The isolated galaxy with a very low central
         color is NGC\,6744, which is clearly devoid of young stars all
         inside its inner ring.}
\label{fig:histo_color}
\end{figure}

Fig.~\ref{fig:histo_color} compares the $F_{15}/F_7$ distributions observed in
the disk and in the central regions of our galaxies (whenever the radius of the
central regions $R_{\rm CNR}$, fitted on 7\,$\mu$m brightness profiles, could
not be defined, the galaxy has been considered as a pure disk). These histograms
indicate that $F_{15}/F_7$ ratios of circumnuclear regions are higher than those
of disks (and this is a systematic property, verified for each individual
galaxy except NGC\,4736 and 6744, whose central regions are dominated by old
stellar populations). Colors of disks are fairly constant and close to the
integrated colors of SA-SAB galaxies ($F_{15}/F_7 = 0.89 \pm 0.14$ for the
$1 \sigma$ dispersion), whereas circumnuclear colors form a broader distribution
extending towards high values ($F_{15}/F_7 = 1.59 \pm 0.78$).

The cause for this difference of colors can easily be seen in the spectra
of Fig.~\ref{fig:spectres}: in all spectra with sufficient signal-to-noise
ratio, the relative intensities
of the UIBs are almost unchanged from galaxy to galaxy, or from central regions
to disks. On the contrary, the level and spectral slope of the continuum seen
longward of 13\,$\mu$m is highly variable and always stronger in the central
regions than in the disks. This continuum is attributed to VSGs (see
Sect.~\ref{sec:sec_obs}) and its presence in the 15\,$\mu$m band is a
characteristic sign of intense star formation \citep[{\it e.g.} ][]{Laurent}.

The reason why high global $F_{15}/F_7$ colors require a high flux concentration
can be directly derived from Fig.~\ref{fig:histo_color}: only the central
regions of galaxies are able to reach high $F_{15}/F_7$ colors, and they have to
dominate the integrated emission to affect the global color. Furthermore,
the fact that the two color histograms overlap explains why a high flux
concentration does not necessarily imply a high $F_{15}/F_7$ color.

\begin{figure}[!t]
\centerline{\resizebox{9.5cm}{!}{\rotatebox{90}{\includegraphics{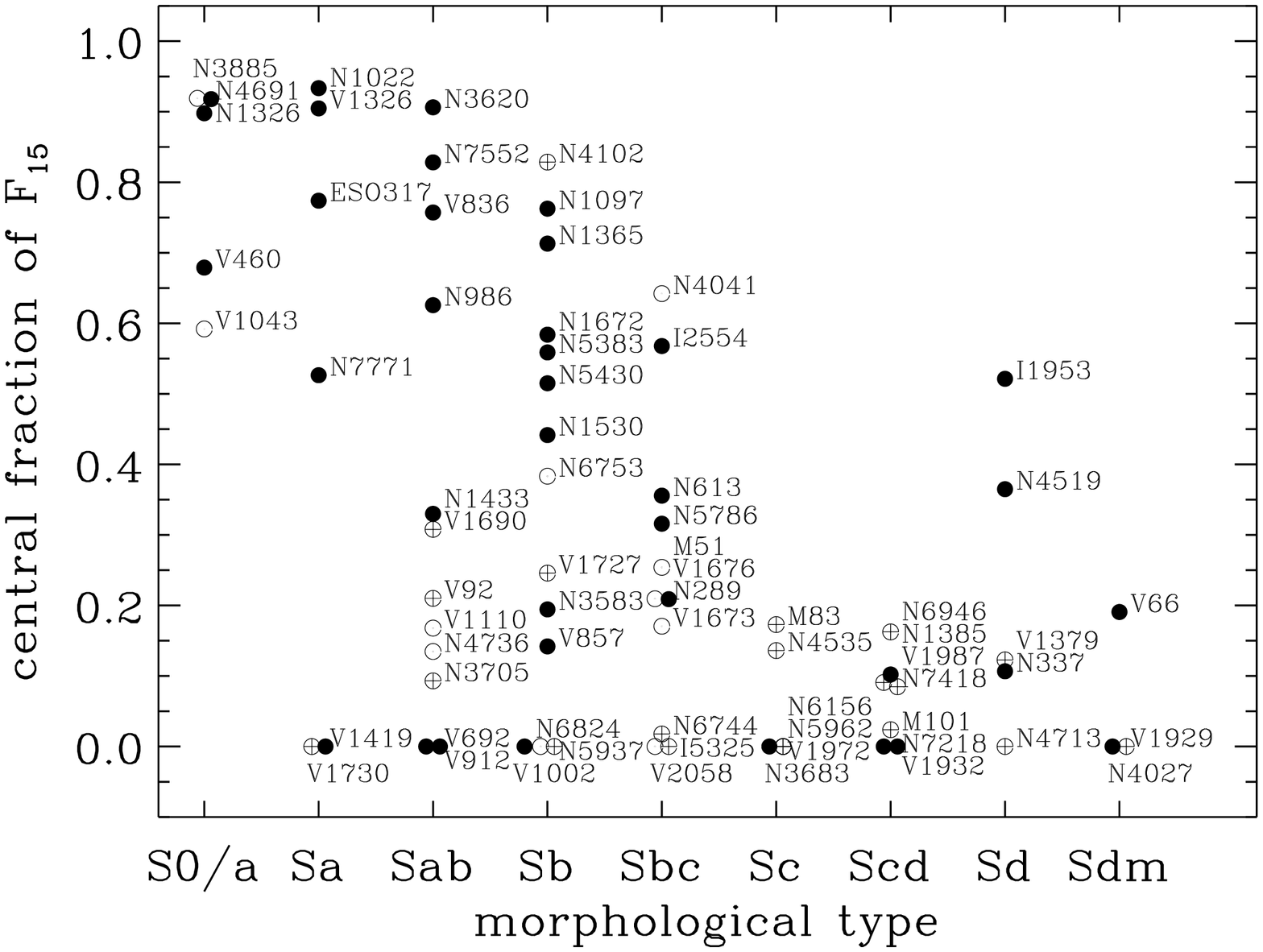}}}}
\caption{Fraction of total 15\,$\mu$m fluxes arising from the central condensation,
         as a function of morphological type. As in Fig.~\ref{fig:cfrac15_color},
	 galaxies with no identifiable central regions have been placed at a null
         ordinate. The central fraction of $F_7$ fluxes, not shown here, has a
         very similar behavior, with only slightly lower values.}
\label{fig:cfrac15_type}
\end{figure}

We however still have to identify the property or properties required,
in addition to belonging to the SB class, for a galaxy to show a high
mid-infrared flux concentration. We have seen in Fig.~\ref{fig:lwcolor_type}
that the morphological type plays a major part in the presence of high
colors. Fig.~\ref{fig:cfrac15_type} shows the evolution of the
concentration fraction as a function of morphological type. It
confirms that for SB galaxies, there is a definite trend for
the central flux fraction to rise as the morphological type gets
earlier. More precisely, SB galaxies with central fractions greater
than 40\% are found predominantly among galaxies earlier than Sb.

It is less clear in Fig.~\ref{fig:cfrac15_type} whether SA-SAB
galaxies follow a similar or a different trend, partly because of the
lack of such bar classes in our sample for types S0/a and Sa, and also
because types Sab and Sb may be incorrect due to the morphological
classification bias affecting cluster galaxies, as already discussed
in Sect.~\ref{sec:lwglobal}. In that section however, we emphasized
the existence of a set of five early-type SA-SAB spirals which do not
suffer from morphological misclassification: VCC\,92 = NGC\,4192, NGC\,3705,
NGC\,4736, NGC\,5937 and NGC\,6824. As apparent in
Fig.~\ref{fig:cfrac15_type}, they all have a low central flux
fraction, much lower than that observed in SB galaxies in the same
range of types. This supports the view that the trend seen for
increasing concentration fraction with earlier type concerns only SB
galaxies (or peculiar objects like VCC\,1043), SA-SAB galaxies having a
generally low concentration factor whatever their type.

We can summarize our findings in this section in the following way:
integrated $F_{15}/F_7$ colors of galaxies are generally of the order of 1.
However, $F_{15}/F_7$ is often higher in central regions.
Spiral galaxies with high $F_{15}/F_7$ colors must simultaneously be (1)
dominated by their central regions, (2) of bar type SB, and (3) of
morphological type earlier than Sb. However, the reverse is not true: as
can be seen in Fig.~\ref{fig:cfrac15_type}, NGC\,5383 (a Markarian galaxy),
1672, 1365 and 1097 for instance fulfill these conditions -- between 55 and
75\% of their 15\,$\mu$m radiation comes from small central regions (respectively
17, 8, 6 and 8\% of the optical diameter) -- yet their $F_{15}/F_7$ color is
very similar to that of disk-dominated galaxies. This suggests that they host
at their center larger concentrations of gas and dust than in the average
of galaxies of the same Hubble type, but for some yet undetermined reasons,
they presently undergo smooth star formation instead of a nuclear starburst.
We propose that either the net gas inflow rate to the center has decreased
(due to a slower replenishment from the inner disk which would have been
previously partially depleted in gas, or a smaller efficiency of the evolved
bar to make gas lose its angular momentum) or, since star formation bursts
occur on a much shorter timescale than bar life, that we are imaging
these objects at a period of quiescence in-between bursts.
Concerning this last point, see the results of the simulations of
\citet{Martinet} and the population synthesis estimates of
\citet{Kotilainen} for the circumnuclear rings of NGC\,1097 and 6574.

\section{Origin of the circumnuclear infrared excess}
\label{sec:orig}

\subsection{Non-stellar activity}
\label{sec:seyfert}

Of the SB galaxies, four are known to host a Seyfert nucleus: in order of
decreasing flux fraction from the central condensation, VCC\,836 = NGC\,4388,
NGC\,1365, NGC\,1097 and NGC\,1433. For these, the high central color could arise
from dust heated by non-stellar radiation from the accretion disk and halo of the
central object and would thus not necessarily indicate the presence of massive
stars. For NGC\,1097, we have the direct visual evidence that the contribution from
the active nucleus to the circumnuclear emission is negligible, since the central
mid-infrared source is resolved into the well-known star-forming ring, which is
very bright, and a faint point source at the nucleus. Correcting the images
of NGC\,1097 for dilution effects with a procedure analog to CLEAN (see the Atlas
for more detail), we obtain fractions of the total circumnuclear fluxes contributed
by the nuclear point source of less than 3\% at 7\,$\mu$m and about 1\% at 15\,$\mu$m.
This central source was measured inside a radius of $3\arcsec$, while the ring
extends between radii $\approx 6\arcsec$ and $12\arcsec$.

We can also inspect the low-resolution spectra between 5 and 16\,$\mu$m of
the central regions of NGC\,1097 and 1365 (left column of
Fig.~\ref{fig:spectres}). Indeed, \citet{Genzel} and \citet{Laurent} have shown
that a strong continuum at 5\,$\mu$m and small equivalent widths of the UIBs are
signatures of dust heated by an active nucleus. Yet all our spectra are
similar to that of the inner plateau of NGC\,5194 ($\approx 50\arcsec$
in diameter) -- which also contains a weak Seyfert nucleus, but completely
negligible -- and to that of NGC\,5236: they are dominated by UIBs in the
5--10\,$\mu$m range and the underlying
continuum at 5\,$\mu$m is comparatively very low. We conclude that in these
galaxies, the contribution of non-stellar heating to the emission observed
inside $R_{\rm CNR}$ is small.

The cases of NGC\,4388 and NGC\,1433 can only be discussed on the basis
of imaging results. The central condensation of NGC\,1433 is large (we
have determined a diameter of $31\arcsec \approx 1.7$\,kpc) and
extremely smooth, much flatter than the point spread function: we
therefore consider unlikely a major contribution from the
{\scriptsize LINER}/Seyfert nucleus, which should manifest itself as a point
source. For NGC\,4388, we cannot conclude and the active nucleus may
be dominant. We can only mention that its global color is lower than that
of VCC\,1326 = NGC\,4491, and this is marginally true as well for the
nucleus, and that the nucleus of VCC\,1326 is not classified as
active\footnote{Far-infrared diagnostics of nuclear activity
are ambiguous for
NGC\,4388: its infrared to radio flux ratio as defined by \citet{Condon}
is $q = 2.27$, and its spectral index between 25 and 60\,$\mu$m
\citep{Grijp} is $\alpha = 1.23$. Both values indicate that
stellar and non-stellar excitations may contribute in comparable
amounts to the infrared energy output. Concerning VCC\,1326, its
spectral index between 25 and 60\,$\mu$m, $\alpha = 1.67$, is rather
typical of starbursts.}.

Hence, the presence of Seyfert nuclei does not modify our
interpretation that high mid-infrared colors in the present sample are
not due to dust heated by non-stellar photons and should rather signal
the existence of central starbursts.

\subsection{Circumnuclear starbursts}

We now examine the most likely cause of the 15\,$\mu$m emission excesses
detected in our sample, central starbursts triggered by the bar
dynamical effects. We warn that NGC\,1022 and NGC\,4691 should be
considered apart: their dust emission comes almost exclusively from
central regions of $\approx 1$\,kpc, but this is more likely due to a
past merger than to the influence of the bar, which may have been
formed or transformed simultaneously as the starburst event was
triggered.

\subsubsection{Available molecular gas}
\label{sec:co}

To see if a significant difference exists between the central
molecular gas content of circumnuclear starburst galaxies and
quiescent ones, we have searched the literature for single-dish
CO(1-0) data in the smallest possible beams. Single-dish data are
better suited to our purpose than interferometric data since the
latter are scarcer and do not collect all the emission from extended
structures. The conversion of CO antenna temperatures to molecular
gas masses is approximate for two main reasons: the H$_2$ mass to CO
luminosity ratio varies with metallicity and physical conditions;
and the derivation of CO fluxes requires the knowledge of the source
structure, because it is coupled to the antenna beam to produce the
observed quantity which is the antenna temperature.

Sensible constraints on the structure of CO emission can be drawn from
that observed in the mid-infrared. As dust is physically
associated with gas, the mid-infrared emission spatial distribution
should follow closely that of the gas, but be modified by the
distribution of the star-forming regions that provide the heating, and
which are likely more concentrated than the gas reservoir. Since
gaussian profiles provide an acceptable description of most infrared
central regions at our angular resolution, we have therefore assumed
that the CO emitting regions are of gaussian shape, with half-power
beam width (HPBW) between one and two times that at 7\,$\mu$m. The 7\,$\mu$m
HPBW were derived by matching gaussian profiles convolved with the
point spread function to the observed 7\,$\mu$m profiles\footnote{In
NGC\,1530, the scales of the molecular gas and infrared concentrations
are of the same order. In NGC\,1022, the source HPBW is estimated to
be $\approx 17\arcsec$, which is $\approx 2.9$ times that of the innermost
infrared regions. However, these are clearly more extended than a
central gaussian and not representative of normal CNRs, since likely
gathered by a merger. For this and other galaxies whose central
regions are clearly structured ({\it i.e.} NGC\,1097 and NGC\,4691), detailed
CO maps where the source is resolved were used. It is also the case
for NGC\,1530, 5236 and 6946.}.

To find the meaning of various antenna temperatures (with various corrections)
and which conventions are used in the literature, the explanations of \citet{Kutner}
and \citet{Downes} were of much help. We converted given temperatures to the $T_R^*$
scale\footnote{which includes corrections for atmospheric attenuation and all
instrumental effects except antenna to source coupling.}. We then attempted a
correction of antenna to source coupling, assuming a gaussian source and a
gaussian
diffraction pattern with angular standard deviations $\theta_S$ and $\theta_B$.
The relationship below follows for the source brightness temperature $T_b$,
which is averaged over the beam in the observation, whereas we want to recover
its intrinsic value over the source extent:
\begin{eqnarray}
   T_R^* \times (\theta_S^2 + \theta_B^2) = T_b \times \theta_S^2.
\nonumber
\end{eqnarray}

Table~\ref{tab:tab_sample} contains the beam width of the observations and the
derived H$_2$ masses for the adopted references. A conversion factor
$f = N(H_2)\, /\, I(CO) = 2.3~10^{24}~ {\rm molecules\,m}^{-2}\,({\rm K\,km\,s}^{-1})^{-1}$
\citep{Strong} has been used to compute the mass as:
\begin{eqnarray}
   M_{H_2}\, /\, (2\, m_H) &=& f \times \int T_b\, dV~ ({\rm K\,km\,s}^{-1}) \times 2\, \pi\, (\theta_S\, D)^2
\nonumber \\
   &~& \times~ (1 - exp(-\frac{1}{2} (\alpha_{\rm CNR}\, /\, \theta_S)^2))
\nonumber
\end{eqnarray}
where $m_H$ is the hydrogen atom mass and $D$ the distance (in m). In the above
formula, we estimate the mass only inside the angular radius $\alpha_{\rm CNR}$
used for the infrared photometry of circumnuclear regions. Only when the central
regions are resolved and mapped is there no need to assume a brightness distribution.
H$_2$ masses derived in this way are probably not more precise than by a factor
three, including the dispersion of the factor $f$, but the dynamic range in the
sample is still sufficient to allow a discussion of the results.

Although the beam of CO observations is in general larger than $\alpha_{\rm CNR}$,
it remains (except for NGC\,337) smaller than the diameter of the bar which
collects gas from inside corotation, believed to be located close to the end of
the bar \citep{Athanassoulab}, so that it is still meaningful to compare our
measurements on infrared condensations to CO data.

\begin{figure}[!t]
\centerline{\resizebox{9.5cm}{!}{\rotatebox{90}{\includegraphics{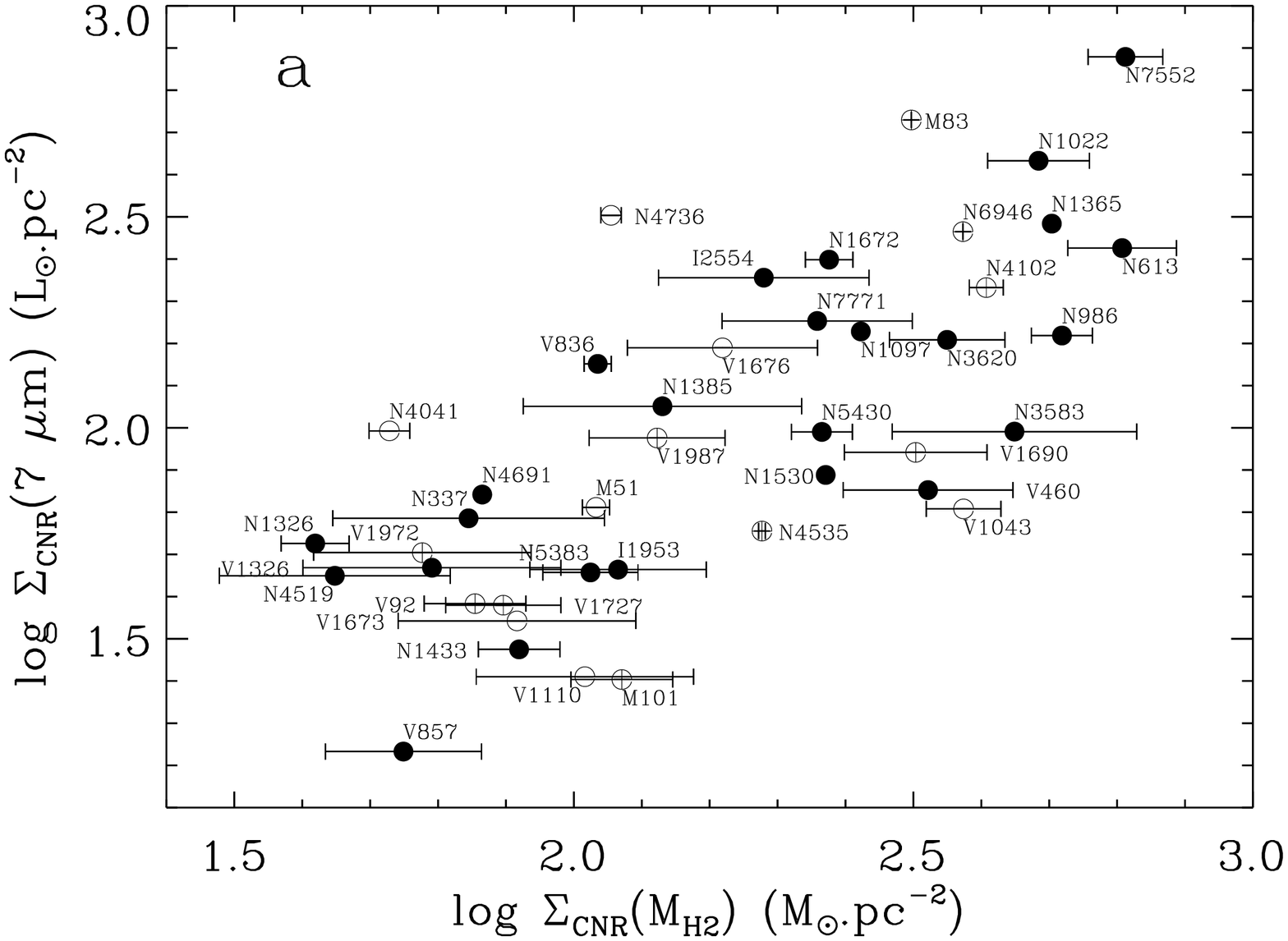}}}}
\centerline{\resizebox{9.5cm}{!}{\rotatebox{90}{\includegraphics{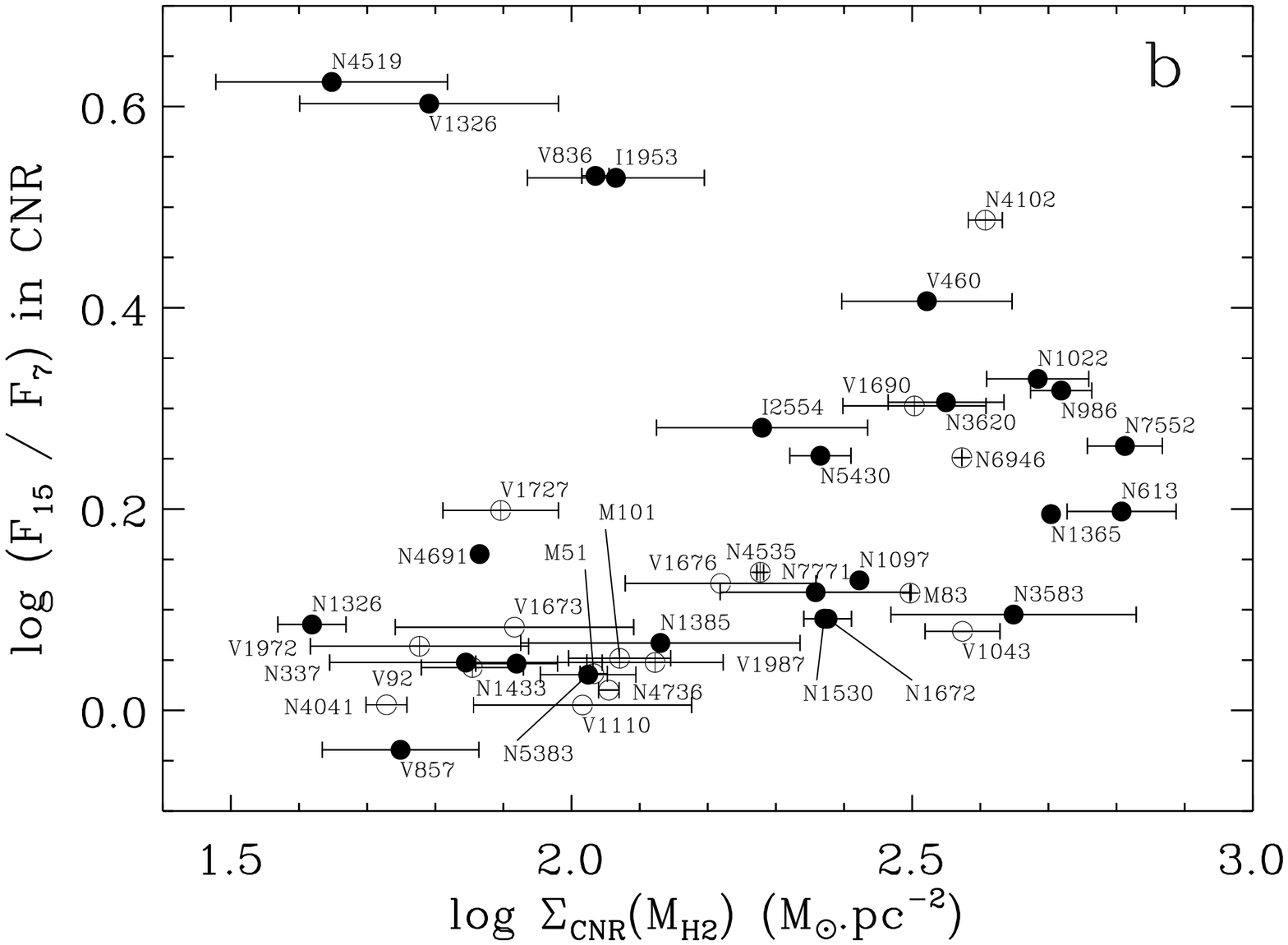}}}}
\caption{{\bf (a):} 7\,$\mu$m surface brightness as a function of the average H$_2$
         surface density, both inside the circumnuclear regions defined by
         mid-infrared photometry (CNR).
         The limits on H$_2$ mass are not true error bars, but simply
         indicate the effect of varying the scale of the gaussian distribution
         from once to twice that measured at 7\,$\mu$m (see text).
         {\bf (b):} $F_{15}/F_7$ color as a function of the average H$_2$ surface
         density, both inside the CNR, with the same convention for error bars
         as in {\bf (a)}.}
\label{fig:co_color}
\end{figure}

Fig.~\ref{fig:co_color}a shows the variation of the 7\,$\mu$m surface brightness
as a function of the average molecular gas surface density inside $R_{\rm CNR}$.
Higher densities of the molecular material are associated with an increase in the
infrared brightness of the central regions. This is expected, since the amount of
dust scales with that of gas, which essentially consists of the molecular phase in
central regions of galaxies. More interesting is Fig.~\ref{fig:co_color}b where
we show the evolution of the $F_{15}/F_7$ color inside $R_{\rm CNR}$ as a function
of the same quantity as in Fig.~\ref{fig:co_color}a. For the majority of our sample,
$F_{15}/F_7$ tends to rise, within a very large dispersion, when the molecular gas
mean density increases (it roughly doubles when the H$_2$ surface brightness
varies by 1.2\,dex). However, a few galaxies dramatically depart from this trend:
for colors higher than 2.5 (log~$F_{15}/F_7 > 0.4$), there is a reversal in the sense
that hot circumnuclear regions seem to be depleted in molecular gas, with respect
to the normal H$_2$ content--color distribution.

Although one can think of several reasons why their molecular content may be
underestimated (the standard conversion factor may not apply for these galaxies
due to their starburst nature or possibly due to a lower metallicity), it is
unlikely that this is the case. First, the implied underestimation factors
appear quite large, at least 4 to 10. Second, if we were to correct the
H$_2$ masses by these factors to bring the galaxies within the trend observed
in Fig.~\ref{fig:co_color}b, then these objects would become abnormal in
Fig.~\ref{fig:co_color}a, with a deficit of 7\,$\mu$m emission\footnote{Note that
it is however conceivable that a fraction of UIB carriers are destroyed, which
would cause such a deficit.}.
The four deviating galaxies do not share a common property which would make them
special with respect to all the others. NGC\,4519 and IC\,1953 are similar SBd
galaxies, VCC\,1326 = NGC\,4491 is a small and low-luminosity SBa,
and VCC\,836 = NGC\,4388 is an edge-on Seyfert SBab (for which the molecular
content may be ill-determined due to the integration of the CO line throughout
the disk).

We thus propose the following interpretation for the galaxies that
wander off the main trend in Fig.~\ref{fig:co_color}b: the main
distribution corresponds to galaxies where the central starburst is more
and more intense, as indicated by the high gas surface densities and colors.
Galaxies at the turnover of the sequence may be observed in a phase of
their starburst (not necessarily common to all galaxies) when it has consumed
or dispersed most of the accumulated gas, because of a higher star formation
efficiency. This suggests an interesting analogy with H{\scriptsize II} regions,
for which the distinction between ``ionization-bounded'' and ``density-bounded''
is made \citep[see][, also for a discussion of the efficiency of molecular
cloud dispersal by young stars]{Whitworth}. Dust should then be depleted too;
however, because of the presence of massive stars, the remaining
dust is exposed to a very intense radiation field and reaches a
high $F_{15}/F_7$ color. This ratio may also increase due to the fact
that the dust which was mixed with rather dense molecular clouds, of
low $F_{15}/F_7$ color, has been dispersed too. Alternatively, the concentrations
of molecular gas in these galaxies may be more compact than in the others
and diluted in our large beam (we cannot exclude that the mid-infrared
distribution includes an unresolved core which dominates the color).
A confirmation of the above scenario clearly requires better measurements of
the central gas content and high-resolution characterization of the starbursts.

Leaving the four galaxies in the upper left quadrant of
Fig.~\ref{fig:co_color}b apart, the data support an interpretation in terms of
starburst with standard properties: the infrared activity in galactic centers can
be stronger when the available molecular gas is denser.

\subsubsection{Color of the central concentration and age of the starburst}

\begin{figure}[!b]
\centerline{\resizebox{9.5cm}{!}{\rotatebox{90}{\includegraphics{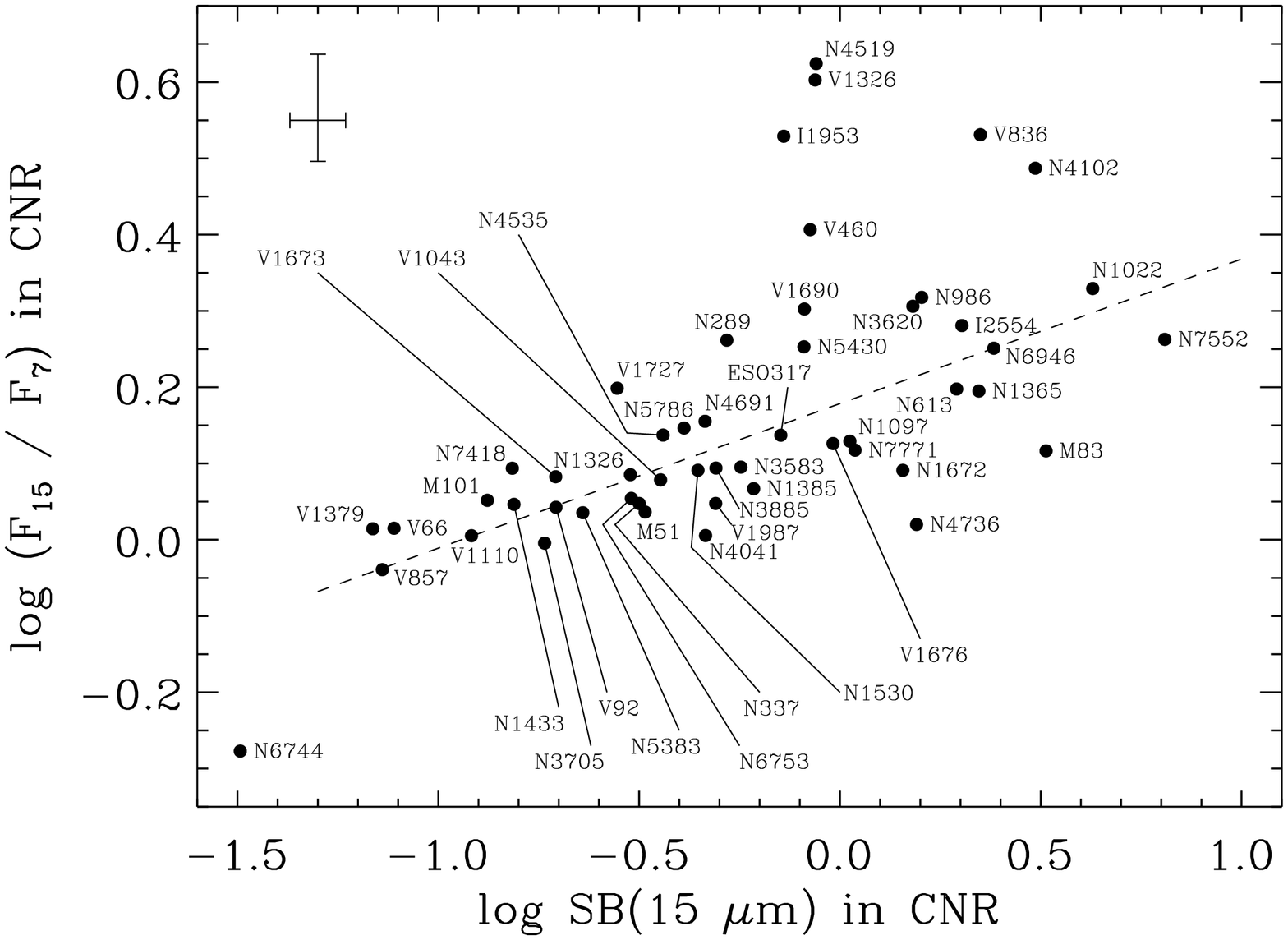}}}}
\caption{Variation of the mid-infrared color with the 15\,$\mu$m surface brightness
         (in mJy\,arcsec$^{-2}$), both inside the same aperture centered on
         circumnuclear regions. The mean error bar is shown in the upper left
         corner. The average location of disks in the diagram in terms of average
         color and global surface brightness (the disk area being delimited by the
	 blue isophote $\mu_{\rm B} = 25$ mag\,arcsec$^{-2}$) would be at
         ($-2.2, -0.05$). The dashed line represents the formal least absolute
         deviation fit, performed including all the galactic central regions
         (used to define the ``color deviation'' in Fig.~\ref{fig:pop}).}
\label{fig:bs_color}
\end{figure}

Fig.~\ref{fig:bs_color} indicates how the $F_{15}/F_7$ color inside $R_{\rm CNR}$
varies with the 15\,$\mu$m surface brightness in the same aperture. In principle,
the mid-infrared surface brightness can increase either because the amount of dust
in the considered area is higher (such as observed in Fig.~\ref{fig:co_color}a),
or because the energy density available to heat the dust increases. The trend
for higher $F_{15}/F_7$ ratios at large 15\,$\mu$m surface brightnesses seen in
Fig.~\ref{fig:bs_color} indicates that indeed, the increase of the 15\,$\mu$m
surface brightness is at least partly due to rise of the mean energy density
in the CNR. In this diagram again, the galaxies with a peculiar behavior in
Fig.~\ref{fig:co_color}b stand apart, well above the locus defined by the least
absolute deviation fit\footnote{We give this name to a fit where
the quantity to be minimized is the sum of absolute values of the distances
between the data points and the line. This method is less sensitive to
outliers than the least squares fit.} (dashed line). This supports the fact that
the trend seen in Fig.~\ref{fig:co_color}b is not due to an underestimation
of the H$_2$ content, and lends further credit to the interpretation presented
in Sect.~\ref{sec:co}.

Another study by \citet{Dale2} has already dealt with the joint variations
of mid-infrared surface brightnesses and colors. However, contrary to
Fig.~\ref{fig:bs_color} where the surface brightnesses and colors are those
of the same physical region (the CNR) in a large sample of galaxies, in the
\citet{Dale2} study, resolution elements inside the target galaxies are first
binned according to their surface brightness before the mean color of the bin
is computed. As a result, a bin does not correspond to a
physical object. We simply note that if galactic central regions are
binned by surface brightness in Fig.~\ref{fig:bs_color}, then the
obtained mean locus is comparable to those shown by \citet{Dale2}.

The galaxies with the highest central $F_{15}/F_7$ colors ($F_{15}/F_7 > 2.5$) and
which stray from the main trend are barred, but their bars are of moderate lengths
(once deprojected and normalized by the optical diameter). In NGC\,4519, NGC\,4102,
VCC\,1326 = NGC\,4491 and IC\,1953, for which it could be estimated,
$D_{\rm bar} / D_{25} \approx 0.2$--0.3, when this ratio ranges between 0.06 and
0.67 in galaxies with measurable bar length. This confirms that the central
activity, signalled by a high $F_{15}/F_7$ color, is not an increasing function of
bar strength, as can be expected from the different timescales for star formation
and bar evolution.

Since the bar strength alone is not sufficient to explain the observed
mid-infrared colors, and since the observational uncertainties are much smaller
than the scatter present in Fig.~\ref{fig:bs_color}, one may suspect that part
of this scatter is due to intrinsic properties of each of the circumnuclear
starbursts considered. Indeed, given that mid-infrared emission likely traces star
formation on timescales longer than, for instance, recombination lines, it is
reasonable to expect that for similar mid-infrared brightnesses (corresponding
to similar gas and energy densities), the mid-infrared color could vary as a
function of the age of the stellar populations responsible for dust excitation.
Since star formation does not happen instantaneously all through a
$\approx 1$\,kpc region and likely occurs in cycles triggered by instabilities,
these stellar populations are multiple and their ages should be weighted to
reflect the successive generations of stars contributing to dust heating.

Using the population synthesis results of \citet{Bonatto}, based on ultraviolet
spectra between 1200 and 3200\,\AA, we can estimate the mean stellar age in the
central $10\arcsec \times 20\arcsec$, weighted by the fraction of luminosity
emitted at 2650\,\AA\ by different population bins. This was possible
for eleven galaxies of our sample in common with the sample of \citet{Bonatto}.
We compare in Fig.~\ref{fig:pop} this mean age to the $F_{15}/F_7$ color
deviation, defined as the difference between the observed color and
that predicted by the mean distribution of all galaxies (indicated by the least
absolute deviation fit in Fig.~\ref{fig:bs_color}) at the same 15\,$\mu$m
surface brightness. For this purpose, we performed the mid-infrared photometry
in a slit aperture identical to that used by \citet{Bonatto}. We indeed see that
the younger the weighted age, the higher the central $F_{15}/F_7$ color deviation.
This is thus in agreement with our hypothesis
that much of the color variations in Fig.~\ref{fig:bs_color} may be due to age
variations of the exciting populations.

There are grounds to think that some scatter in Fig.~\ref{fig:pop} is due to the
methodology adopted by \citet{Bonatto} in their study. They have grouped galaxies
of their sample according to spectral resemblance, morphological type and
luminosity, and co-added all UV spectra of each group in order to increase the
signal to noise ratio before performing the population synthesis. However, it may
not be fully justified to average spectra of different galaxies with the same
overall shape but different spectral signatures. A further drawback of this
study is that it cannot properly take into account extinction, because of
the limitation
to a small spectral range in the UV: the derived very low extinctions are
meaningless. That is why the two galaxies departing from the well-defined trend
described above could owe their age to the method rather than to their
intrinsic properties:

\begin{figure}[!t]
\centerline{\resizebox{9.5cm}{!}{\rotatebox{90}{\includegraphics{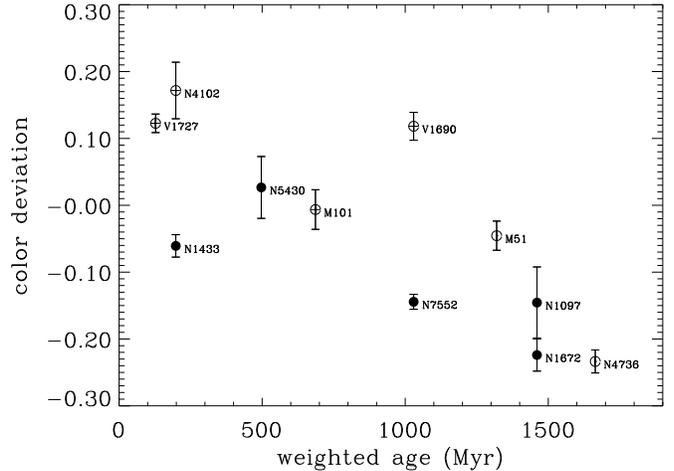}}}}
\caption{The abscissa indicates the mean age of stellar populations, according
         to the synthesis results of \citet{Bonatto}, including the first six
         elements of their base, stellar clusters to which they attribute ages
         between 0 and 0.7\,Gyr for the first five and in the interval 0.7-7\,Gyr
         for the last one, but excluding the oldest element, an elliptical bulge
         representing ages between 7 and 17\,Gyr. The ages are weighted by the
         fraction of the flux at 2650\,\AA\ that each different population emits.
         The contribution from continuous
         star formation has been approximated by a constant flux fraction
         (equal to the minimum value) and subtracted, in order to consider only
         successions of bursts. The ordinate is the difference between the
         measured $F_{15}/F_7$ color and the color expected from the mean
         relationship between central surface brightnesses and colors shown in
         Fig.~\ref{fig:bs_color}. For this graph, the photometry was performed
         inside the same apertures as in \citet{Bonatto},
         $10\arcsec \times 20\arcsec$, and error bars show the effect of
         varying the slit orientation, which is not given by \citet{Bonatto}.}
\label{fig:pop}
\end{figure}

\begin{list}{--}{}

   \item VCC\,1690 = NGC\,4569 is assigned the same large weighted age
   as NGC\,7552 (their UV spectra are co-added), but has a higher color
   excess with respect to the mean distribution in
   Fig.~\ref{fig:bs_color}. In fact, \citet{Maoz} detected very
   strong P~Cygni absorption lines of high-excitation ions
   (C{\scriptsize IV}, Si{\scriptsize IV}, N{\scriptsize V})
   characteristic of the winds of massive young stars ($< 6$\,Myr), and
   its spectrum between 1220 and 1590\,\AA\ is nearly identical to that
   of the starburst NGC\,1741B.

   \item NGC\,1433 is co-added with NGC\,4102, which results in
   a small weighted age. Yet its color excess is much lower than that of
   NGC\,4102 and more comparable to that of NGC\,7552. There is some
   evidence that the extinction in the central regions of NGC\,1433 is
   much lower than in other galaxies: whereas available Balmer
   decrement measures indicate H$\alpha$ absorptions of the order of 2--3
   in other nuclei, the decrement given by \citet{Diaz} for NGC\,1433 indicates
   A(H$\alpha$)~$\approx 0.9$. Even if the Balmer decrement is not a good
   extinction measure, it is instructive to compare values in
   different galaxies. We also notice that NGC\,1433 is the only one
   among strongly barred spirals which has an amorphous circumnuclear
   region, with no hot spot that would indicate the presence of
   massive stellar clusters.

\end{list}

To conclude, Fig.~\ref{fig:bs_color} is a strong indication that the
mid-infrared emission in circumnuclear regions is influenced by successive
episodes of star formation over relatively long periods of time: on
the mean, the $F_{15}/F_7$ color is a sensitive function of the mid-infrared
surface brightness, but this relationship is modulated by the mean age of
the stellar populations. A strinking example of this is NGC\,4736. Its
central mid-infrared brightness is in the high range,
but its central $F_{15}/F_7$ ratio is low, in accordance with its
large mean stellar age confirmed by \citet{Taniguchi}.
From optical population synthesis, they find that a
central starburst occurred about 1\,Gyr ago in this galaxy, and that
subsequent nuclear star formation has proceeded at a low rate.

Combining this result with that of Sect.~\ref{sec:co},
we can form the following sketch of what determines the
mid-infrared properties of circumnuclear regions: the central
surface brightness is connected to the amount of gas, as expected
if gas-to-dust ratios are relatively constant. However,
accumulation of gas in the center allows the triggering of intense star
formation, so that the interstellar radiation field
increases, reflected in higher $F_{15}/F_7$ ratios. Fig.~\ref{fig:bs_color}
and \ref{fig:pop} suggest then that deviations from this simple
description can be related to the star formation history of the
circumnuclear regions. On-going starbursts produce excess $F_{15}/F_7$ colors,
while faded starbursts are associated with $F_{15}/F_7$ deficits.

Additional variation in mid-infrared colors may
arise from differences in metallicity and in the compactness of the
starburst, with consequences on the amount and nature of the dust,
but this is out of the scope of the present study.

\section{Summary and conclusions}

We have studied the mid-infrared activity induced by bars in a sample of
69 nearby spiral galaxies with infrared luminosities spanning a large
range below the class of luminous infrared galaxies.
We have found that:

\begin{list}{--}{}

\item The mid-infrared emission of the normal galaxies in our sample is
    essentially contributed by a thermal continuum from very small grains
    (VSGs) longward of 10\,$\mu$m and the family of aromatic bands (UIBs)
    detected in a wide diversity of environments. It is the variation
    of the VSG component with respect to the UIBs that is responsible for
    $F_{15}/F_7$ changes in our galaxies.
    From the comparison with observations of resolved Galactic regions,
    this can be related to an increase of the filling factor of star forming
    complexes by photoionized regions, hence a decrease of the contribution
    to the mid-infrared emission from neutral and molecular media.

\item There is a dichotomy between spiral disks, where the
    integrated $F_{15}/F_7$ color is close to 1 and shows little
    dispersion, and circumnuclear regions, where $F_{15}/F_7$ ranges from
    disk-like to high values (up to 4). We have found no indication
    that destruction of UIB carriers occurs at the scale of
    circumnuclear regions, although it would be desirable to analyze
    infrared spectroscopic data of the most active galaxies, to
    elaborate on this.

\item We confirm that barred spirals distinguish themselves from
    unbarred galaxies in the sense that they can reach higher $F_{15}/F_7$
    colors. This effect is however restricted to early
    morphological types, in agreement with previous IRAS-based studies
    \citep{Hawarden, Huang}. We show unambiguously that this
    emission excess arises in circumnuclear regions which can completely
    dominate the mid-infrared emission, although their size remains
    modest ($D_{\, \rm CNR}$ ranges between 2 and 26\% of the optical diameter
    in the whole sample, with no clear dependency on Hubble type).
    Galaxies with a global color excess are all dominated by their
    central regions. This is a confirmation of predictions from
    hydrodynamical models \citep{Athanassoulab, Friedli},
    according to which a barred perturbation, through tidal torques and
    shocks, induces substantial mass transfer towards circumnuclear
    regions. We observe the consequences of these gas flows, {\it i.e.} the
    intense star formation that they fuel.

\item An important fact to mention is that only a fraction of
    early-type barred galaxies can be distinguished from unbarred
    galaxies in their infrared properties. Several interrelated
    parameters may explain this quiescence of many barred galaxies.
    With the present data we are unable to address this issue and thus
    only list possible explanations: a bar evolves on a much longer
    timescale than a starburst \citep{Martinet} and the accretion
    rate by a bar is slow; the inward mass transfer is regulated by the
    depth of the potential well, the intensity of the shocks inside the
    bar, the star formation efficiency along the path of the inflowing
    gas before it reaches central regions \citep{Martina}, etc.

\end{list}

Although the presence of a bar can be an efficient means
of triggering
circumnuclear starbursts, the dust emission processes in central
regions are the same in barred and unbarred galaxies. We have studied
the properties of these central regions at the
degree of detail accessible to our spatial resolution. Several physical
properties were found to control the mid-infrared color $F_{15}/F_7$.

\begin{list}{--}{}
\item The estimated molecular gas content inside the central
    regions ($R_{\rm CNR}$): as
    expected, the mid-infrared brightness at 7\,$\mu$m tends to rise with
    increasing mean gas density, which reflects the physical
    association of dust with molecular material. The $F_{15}/F_7$
    color is also correlated with the mean gas density. As a higher
    gas content allows more efficient star formation on relatively large
    scales, according to the Schmidt law and stability criteria
    or other empirical laws \citep{Kennicutt}, this
    supports an interpretation of mid-infrared colors in terms of
    starburst intensity. A few galaxies, showing the most extreme
    $F_{15}/F_7$ ratios, depart from the general trend. This could be
    interpreted as a transient evolutionary state of the starburst
    during which most of the gas has been consumed or dispersed,
    although a definite assessment of the effect requires better
    molecular line data.

\item The age of the stellar populations heating dust: dust is
    sensitive to star formation on relatively long timescales, as can
    be expected from the fact that it can be heated by optical and near-UV
    photons. However, the strength of the VSG continuum is more
    sensitive to the radiation energy density and hardness than UIBs, and the
    results that we report here indicate that $F_{15}/F_7$ excesses are
    linked to the weighted age of the exciting stellar populations.
    Mid-infrared colors are therefore influenced by the previous star
    formation history over at least 1\,Gyr, and depend on the fraction of the
    ultraviolet radiation power contributed by young populations created
    in a contemporary starburst, with respect to intermediate-age populations
    from already faded bursts.
\end{list}

\begin{acknowledgements}
We thank our referee, Louis Martinet, for his helpful remarks.
\vspace*{1ex} \\
The ISOCAM data presented in this paper were analyzed using and adapting
the CIA package, a joint development by the ESA Astrophysics Division and
the ISOCAM Consortium (led by the PI C. Cesarsky, Direction des Sciences
de la Mati\`ere, C.E.A., France).
\end{acknowledgements}

\bibliographystyle{apj}

\clearpage

\input{ms1024_tab1.tex}
\input{ms1024_tab2.tex}

\end{document}

%% file: ms1024_tab1.tex
\begin{table*}[!t]
\caption[]{General properties of sample galaxies. Virgo members are also named
           from the VCC catalog. Other galaxies belong to the field or loose
           groups, unless otherwise noted. Distances are from the NGC catalog
           of \citet{Tully}, taking into account the Virgo infall and assuming
           $h_{100} = 0.75$; morphological types, asymptotic blue magnitudes
           $m_{\rm BT}$ and major diameters $D_{25}$ at the isophote
           $\mu_{\rm B} = 25\,{\rm mag\,arcsec}^{-2}$ are from the RC3
           \citep{Vaucouleurs}.}
\label{tab:tab_sample}
\begin{minipage}[t]{18cm}
\begin{flushleft}
\begin{tabular}{|l@{\hspace*{0.1cm}}c@{\hspace*{0.1cm}}rl@{\hspace*{0.1cm}}rrl@{\hspace*{0.3cm}}ll@{\hspace*{0.1cm}}l|}
\hline
\noalign{\smallskip}
name & RA DEC (2000) & $D$~~ & morph. & $m_{\rm BT}$ & $D_{25}$ &
H{\scriptsize I} def.$^a$ & $M_{\rm H2}$ ~~~~~-- ($b$)$^b$ & nuclear &
\hspace*{-1em} tidal$^c$ \\
~ & ~ & (Mpc) & ~~type & ~ & ($\arcmin$)~ & ~ & (log\,M$_{\odot}$) -- (\arcsec) &
~~type & ~int. \\
\noalign{\smallskip}
\hline
\noalign{\smallskip}
N289            & 00 52 42$-$31 12.4 & 19.4 & SBbc  & 11.72 &  5.13 & -1.75 (P)     & ~ 	            & L/H{\scriptsize II} (V) & ~ \\
N337            & 00 59 50$-$07 34.7 & 20.7 & SBd   & 12.06 &  2.88 & -0.82 (T)     & 8.05/7.65 (44) (E)    & H{\scriptsize II} (V)   & mbs \\
N613            & 01 34 18$-$29 25.0 & 17.5 & SBbc  & 10.73 &  5.50 & \,\,0.64 (MF) & 9.20/9.04 (44) (E)    & L/H{\scriptsize II} (V) & ~ \\
N1022           & 02 38 33$-$06 40.7 & 18.5 & SBa   & 12.09 &  2.40 & \,\,0.98 (TB) & 8.91/8.76 (44) (GD)   & H{\scriptsize II} (A)   & am \\
N1097           & 02 46 19$-$30 16.4 & 14.5 & SBb   & 10.23 &  9.33 & \,\,0.13 (MF) & ~~~~9.33~~~~(17) (GN) & L/Sy (V/M)              & doa \\
N1365 \dag      & 03 33 37$-$36 08.3 & 16.9 & SBb   & 10.32 & 11.22 & \,\,0.22 (J)  & 9.69/9.68 (44) (SJ)   & Sy1/2 (V/M)             & ~ \\
N1433 \ddag     & 03 42 01$-$47 13.3 & 11.6 & SBab  & 10.70 &  6.46 & \,\,0.87 (RB) & 8.36/8.24 (44) (BW)   & L/Sy2 (V/M)             & ~ \\
N1530           & 04 23 29+75 17.8   & 36.6 & SBb   & 12.25 &  4.57 & -0.10 (RT)    & ~~~~9.65~~~(5.3) (RD) & ~                       & ~ \\
N1672 \ddag     & 04 45 42$-$59 15.0 & 14.5 & SBb   & 10.28 &  6.61 & -1.24 (MP)    & 9.02/8.95 (44) (BW)   & L/H{\scriptsize II} (V) & ~ \\
N4027           & 11 59 30$-$19 16.1 & 25.6 & SBdm  & 11.66 &  3.16 & -0.62 (RH)    & ~                     & H{\scriptsize II} (V)   & mbs \\
N4535 (V1555)   & 12 34 20+08 11.9   & 16.8 & SABc  & 10.59 &  7.08 & \,\,0.88 (GR) & 8.73/8.72 (21) (*)    & H{\scriptsize II} (V)   & ~ \\
N4691           & 12 48 14$-$03 20.0 & 22.5 & SB0/a & 11.66 &  2.82 & \,\,0.24 (HS) & ~~~~9.14~~~(21) (W1)  & H{\scriptsize II} (K)   & am \\
N4736 (M94)     & 12 50 54+41 07.2   &  4.3 & SAab  &  8.99 & 11.22 & \,\,0.77 (HS) & 7.27/7.24 (16) (ST)   & L (H)                   & ~ \\
N5194 (M51)     & 13 29 53+47 11.8   &  7.7 & SAbc  &  8.96 & 11.22 & -0.01 (AF)    & 8.95/8.99 (55) (HB)   & Sy2 (H)                 & doa \\
N5236 (M83)     & 13 37 00$-$29 52.1 &  4.7 & SABc  &  8.20 & 12.88 & -2.40 (HB)    & ~~~~8.24~~~(16) (HN)  & H{\scriptsize II} (V)   & ~ \\
N5383 (Mrk281)  & 13 57 05+41 50.7   & 37.8 & SBb   & 12.05 &  3.16 & \,\,0.36 (S)  & 9.53/9.39 (55) (SR)   & H{\scriptsize II} (A)   & ~ \\
N5457 (M101)    & 14 03 13+54 20.9   &  5.4 & SABcd &  8.31 & 28.84 & \,\,0.58 (HB) & 7.97/7.82 (55) (SI)   & H{\scriptsize II} (H)   & doa \\
N6744           & 19 09 45$-$63 51.4 & 10.4 & SABbc &  9.14 & 19.95 & -0.18 (MF)    & ~                     & L (V)                   & ~ \\
N7552 \S        & 23 16 11$-$42 35.0 & 19.5 & SBab  & 11.25 &  3.39 & -0.44 (T)     & 9.37/9.26 (44) (CS)   & H{\scriptsize II} (V)   & ~ \\
\multicolumn{10}{|l|}{~} \\
\multicolumn{10}{|l|}{\it Virgo cluster sample:} \\
\multicolumn{10}{|l|}{~} \\
N4178 (V66)     & 12 12 46+10 51.8   & 16.8 & SBdm  & 11.90 &  5.13 & \,\,0.33 (GR)    & ~                   & H{\scriptsize II} (H)   & ~ \\
N4192 (V92)     & 12 13 48+14 53.7   & 16.8 & SABab & 10.95 &  9.77 & \,\,0.52 (GR)    & 8.58/8.43 (45) (KY) & L/H{\scriptsize II} (H) & ~ \\
N4293 (V460)    & 12 21 13+18 23.0   & 17.0 & SB0/a & 11.26 &  5.62 & \,\,1.23 (GR)    & 8.66/8.41 (45) (KY) & L (K)                   & ~ \\
N4351 (V692)    & 12 24 02+12 12.4   & 16.8 & SBab  & 13.03 &  2.00 & \,\,0.41 (GR)    & ~                   & ~                       & note \\
N4388 (V836) \# & 12 25 47+12 39.7   & 16.8 & SBab/ & 11.76 &  5.62 & \,\,1.84 (GR)    & 8.41/8.37 (16) (V)  & Sy2 (K)                 & ~ \\
N4394 (V857)    & 12 25 56+18 12.9   & 16.8 & SBb   & 11.73 &  3.63 & \,\,1.79 (GR)    & 8.12/7.89 (45) (KY) & L (K)                   & ~ \\
N4413 (V912)    & 12 26 32+12 36.6   & 16.8 & SBab  & 12.25 &  2.34 & \,\,1.48 (GR)    & ~                   & ~                       & ~ \\
N4430 (V1002)   & 12 27 27+06 15.8   & 16.8 & SBb   & 12.79 &  2.29 & \,\,1.02 (HH)    & ~                   & ~                       & doa \\
N4438 (V1043)   & 12 27 46+13 00.6   & 16.8 & SA0/a & 11.02 &  8.51 & \,\,2.53 (GR)    & 8.99/8.88 (21) (CD) & L (K)                   & pc \\
N4450 (V1110)   & 12 28 29+17 05.1   & 16.8 & SAab  & 10.90 &  5.25 & \,\,1.78 (GR)    & 8.37/8.05 (44) (Bo) & L (K)                   & ~ \\
N4491 (V1326)   & 12 30 57+11 29.0   & 16.8 & SBa   & 13.50 &  1.70 & $\!\!>$1.73 (GK) & 7.73/7.35 (33) (Bo) & ~                       & ~ \\
N4498 (V1379)   & 12 31 40+16 51.2   & 16.8 & SABd  & 12.79 &  2.95 & \,\,1.68 (GR)    & ~                   & ~                       & ~ \\
N4506 (V1419)   & 12 32 11+13 25.3   & 16.8 & SBa   & 13.63 &  1.62 & $\!\!>$1.60 (HG) & ~                   & ~                       & ~ \\
N4567 (V1673)   & 12 36 33+11 15.5   & 16.8 & SAbc  & 12.06 &  2.95 & $\!\!|<$1.04 (C) & 8.32/7.97 (45) (KY) & H{\scriptsize II} (H)   & ~ \\
N4568 (V1676)   & 12 36 35+11 14.3   & 16.8 & SAbc  & 11.68 &  4.57 & $\!\!|$	       & 8.56/8.28 (33) (E)  & H{\scriptsize II} (K)   & ~ \\
N4569 (V1690)   & 12 36 50+13 09.8   & 16.8 & SABab & 10.26 &  9.55 & \,\,2.11 (GR)    & 8.98/8.77 (33) (Bo) & L/H{\scriptsize II} (H) & ~ \\
N4579 (V1727)   & 12 37 44+11 49.2   & 16.8 & SABb  & 10.48 &  5.89 & \,\,1.94 (GR)    & 8.54/8.37 (33) (E)  & L/Sy2 (H)               & ~ \\
N4580 (V1730)   & 12 37 48+05 22.2   & 16.8 & SABa  & 11.83 &  2.09 & \,\,1.92 (HH)    & ~                   & ~                       & ~ \\
N4633 (V1929)   & 12 42 37+14 21.4   & 16.8 & SABdm & 13.75 &  2.14 & \,\,0.40 (HH)    & ~                   & ~                       & ~ \\
N4634 (V1932)   & 12 42 40+14 17.8   & 16.8 & SBcd/ & 13.16 &  2.57 & \,\,1.05 (HH)    & ~                   & ~                       & ~ \\
N4647 (V1972)   & 12 43 32+11 34.9   & 16.8 & SABc  & 11.94 &  2.88 & \,\,1.82 (GR)    & 8.11/7.79 (45) (KY) & H{\scriptsize II} (K)   & ~ \\
N4654 (V1987)   & 12 43 57+13 07.6   & 16.8 & SABcd & 11.10 &  4.90 & \,\,0.16 (GR)    & 8.32/8.12 (21) (Br) & H{\scriptsize II} (V)   & doa \\
N4689 (V2058)   & 12 47 46+13 45.8   & 16.8 & SAbc  & 11.60 &  4.27 & \,\,1.44 (GR)    & ~                   & H{\scriptsize II} (K)   & ~ \\
\noalign{\smallskip}
\hline
\end{tabular}
\end{flushleft}
\end{minipage}
\end{table*}
\setcounter{table}{0}
\begin{table*}[!t]
\caption[]{continued. The last column indicates the name of the observation project:
           {\it S} for {\it Sf\_glx} or {\it H{\scriptsize I}\_Q\_gal} for IC\,1953
           (PI: G. Helou), and {\it I} for {\it Irgal} (PI: T. Onaka).}
\begin{minipage}[t]{18cm}
\begin{flushleft}
\begin{tabular}{|l@{\hspace*{0.1cm}}c@{\hspace*{0.1cm}}rl@{\hspace*{0.1cm}}rrl@{\hspace*{0.3cm}}ll@{\hspace*{0.1cm}}l@{\hspace*{0.1cm}}r|}
\hline
\noalign{\smallskip}
name & RA DEC (2000) & $D$~~ & morph. & $m_{\rm BT}$ & $D_{25}$ &
H{\scriptsize I} def.\footnote{The H{\scriptsize I} deficiency according to the
   definition and reference values of \citet{Guiderdoni}. It is normalized by the
   dispersion in the field sample. Diameters are taken from the RC2 for consistency,
   and H{\scriptsize I} fluxes from the indicated references. NGC\,4567/68 are
   unresolved in H{\scriptsize I}. When $D_{\rm HI}/D_{\rm opt} \leq 1.$ and
   $Def \geq 0.5$ in \citet{Cayatte}, it corresponds to $def > 1.2$ here.} &
$M_{\rm H2}$ ~~~~~-- ($b$)\footnote{The given range represents the effect of varying
   the scale length of the CO distribution from once to twice that of infrared
   circumnuclear regions (see text). $b$ is the beam HPBW of the observations used.
   (*) NGC 4535 was observed by A. Bosma, D. Reynaud and H. Roussel at the IRAM
   30\,m telescope.} &
nuclear & \hspace*{-1em} tidal\footnote{Signs of tidal interaction. ``doa'':
   asymmetrical distortion of outer arms. ``mbs'': magellanic barred spiral.
   ``am'': amorphous. ``pc'': past collision. ``note'': On DSS images, the brightness
   peak is displaced by $\approx 10\arcsec$ ENE from the center of the regular outer
   isophotes and the NW outer disk seems depressed in stars and gas. \\
   \hspace*{1ex} \dag\ members of the Fornax cluster of galaxies.
   \ddag\ members of the Dorado group of galaxies.
   \S\ member of the Grus quartet with NGC\,7582/90/99.
   $\natural$ member of the spiral-rich Ursa Major cluster. \\
   \hspace*{1ex} \#\ classified SA in the RC3, here considered a SB after the
   morphological arguments of \citet{Phillips} and \citet{Mcleod}, and the recent
   kinematic analysis of \citet{Veilleux}. \\
   \hspace*{1ex} $^d$ The distances of ESO\,317-G023, NGC\,6753 and 6156 were assumed
   to be those of the galaxy groups LGG\,199, LGG\,426 and LGG\,407 \citep{GarciaD};
   those of NGC\,5786 and 7771 were estimated from the H{\scriptsize I} redshift,
   that of NGC\,3620 from the CO redshift and those of NGC\,5430, 5937 and 6824
   from the optical redshift (with $h_{100} = 0.75$).} & ~ \\
~ & ~ & (Mpc) & ~~type & ~ & ($\arcmin$)~ & ~ & (log\,M$_{\odot}$) -- (\arcsec) &
~~type & ~int. & ~ \\
\noalign{\smallskip}
\hline
\noalign{\smallskip}
\multicolumn{11}{|l|}{\it Supplementary galaxies:} \\
\multicolumn{11}{|l|}{~} \\
N986                 & 02 33 34$-$39 02.6 & 23.2 & SBab  & 11.64 &  3.89 & \,\,0.47 (RM) & 9.48/9.39 (44) (E)  & H{\scriptsize II} (V)   & ~   & {\it S} \\
N1326 \dag           & 03 23 56$-$36 27.9 & 16.9 & SB0+  & 11.37 &  3.89 & -0.18 (HC)    & 8.43/8.33 (44) (W2) & L (MC)                  & ~   & {\it S} \\
N1385                & 03 37 28$-$24 30.1 & 17.5 & SBcd  & 11.45 &  3.39 & -0.14 (AH)    & 8.31/7.90 (44) (AC) & H{\scriptsize II} (V)   & mbs & {\it S} \\
N3583                & 11 14 11+48 19.1   & 34.0 & SBb   & 11.90 &  2.82 & -0.31 (T)     & 9.45/9.09 (55) (SS) & H{\scriptsize II} (H)   & ~   & {\it S} \\
N3620~$^d$           & 11 16 04$-$76 12.9 & 23.7 & SBab  & ~     &  2.75 & \,\,0.60 (RM) & 9.61/9.44 (44) (E)  & ~                       & ~   & {\it S} \\
N3683                & 11 27 32+56 52.6   & 28.4 & SBc   & 13.15 &  1.86 & -1.46 (Ka)    & ~                   & ~                       & ~   & {\it S} \\
N3705                & 11 30 07+09 16.6   & 17.0 & SABab & 11.86 &  4.90 & -0.28 (HZ)    & ~                   & L/H{\scriptsize II} (H) & ~   & {\it S} \\
N3885                & 11 46 47$-$27 55.4 & 27.8 & SA0/a & 11.89 &  2.40 & -1.08 (C2)    & ~                   & ~                       & ~   & {\it S} \\
N4041                & 12 02 12+62 08.2   & 22.7 & SAbc  & 11.88 &  2.69 & -0.82 (T)     & 8.86/8.80 (33) (E)  & H{\scriptsize II} (H)   & ~   & {\it I} \\
N4102 $\natural$     & 12 06 23+52 42.7   & 17.0 & SABb  & 11.99 &  3.02 & \,\,0.53 (T)  & 9.13/9.08 (21) (M)  & H{\scriptsize II} (H)   & ~   & {\it S} \\
N4519 (V1508)        & 12 33 30+08 39.3   & 16.8 & SBd   & 12.34 &  3.16 & -0.32 (GR)    & 7.63/7.29 (45) (Y)  & ~                       & ~   & {\it S} \\
N4713                & 12 49 58+05 18.7   & 17.9 & SABd  & 12.19 &  2.69 & -0.82 (HZ)    & ~                   & L/H{\scriptsize II} (H) & ~   & {\it S} \\
N5430 (Mrk799)~$^d$  & 14 00 46+59 19.7   & 40.4 & SBb   & 12.72 &  2.19 & \,\,0.81 (TB) & 9.52/9.43 (21) (KS) & H{\scriptsize II} (K2)  & ~   & {\it I} \\
N5786~$^d$           & 14 58 57$-$42 00.8 & 39.7 & SBbc  & 12.17 &  2.34 & -0.96 (RM)    & ~                   & ~                       & ~   & {\it S} \\
N5937~$^d$           & 15 30 46$-$02 49.8 & 36.8 & SABb  & 13.11 &  1.86 & \,\,0.60 (TB) & ~                   & L/H{\scriptsize II} (A) & ~   & {\it I} \\
N5962                & 15 36 32+16 36.5   & 31.8 & SAc   & 11.98 &  2.95 & \,\,0.57 (Kr) & ~                   & H{\scriptsize II} (V)   & ~   & {\it S} \\
N6156~$^d$           & 16 34 52$-$60 37.1 & 42.9 & SABc  & 12.30 &  1.58 & ~             & ~                   & ~                       & ~   & {\it I} \\
N6753~$^d$           & 19 11 23$-$57 02.9 & 42.4 & SAb   & 11.97 &  2.45 & ~             & ~                   & ~                       & ~   & {\it S} \\
N6824~$^d$           & 19 43 41+56 06.6   & 44.5 & SAb   & 13.00 &  1.70 & \,\,0.80 (Ka) & ~                   & ~                       & ~   & {\it I} \\
N6946                & 20 34 52+60 09.2   &  5.5 & SABcd &  9.61 & 11.48 & -1.22 (RS)    & ~                   & H{\scriptsize II} (H)   & ~   & {\it S} \\
N7218                & 22 10 12$-$16 39.6 & 22.0 & SBcd  & 12.70 &  2.51 & -0.70 (T)     & ~                   & ~                       & ~   & {\it S} \\
N7418                & 22 56 36$-$37 01.8 & 17.8 & SABcd & 11.65 &  3.55 & \,\,0.28 (T)  & ~                   & ~                       & ~   & {\it S} \\
N7771 (Mrk9006)~$^d$ & 23 51 25+20 06.7   & 57.4 & SBa   & 13.08 &  2.51 & \,\,0.28 (MS) & 9.86/9.58 (45) (Y)  & H{\scriptsize II} (VK)  & ~   & {\it S} \\
IC1953               & 03 33 42$-$21 28.8 & 22.1 & SBd   & 12.24 &  2.75 & \,\,1.34 (T)  & 8.46/8.20 (44) (CP) & ~                       & ~   & {\it S} \\
IC2554               & 10 08 50$-$67 01.8 & 16.7 & SBbc  & 12.51 &  3.09 & -0.35 (RC3)   & 8.65/8.34 (44) (AB) & ~                       & doa & {\it I} \\
IC5325               & 23 28 43$-$41 20.0 & 18.1 & SABbc & 11.83 &  2.75 & \,\,0.79 (T)  & ~                   & ~                       & ~   & {\it S} \\
ESO317-G023~$^d$     & 10 24 43$-$39 18.4 & 32.8 & SBa   & 13.93 &  1.91 & ~             & ~                   & ~                       & ~   & {\it S} \\
\noalign{\smallskip}
\hline
\end{tabular}
\end{flushleft}
\end{minipage}
\end{table*}

%% file: ms1024_tab2.tex
\begin{table*}[!t]
\caption[]{Photometric results at 15 and 7\,$\mu$m, obtained as described in the
           Atlas (total fluxes, diameter aperture used for central regions, fluxes
           inside this aperture and background levels). We warn the reader that
           the uncertainties can only be taken as order-of-magnitude values (see
           the Atlas), especially for galaxies of the third subsample belonging to
           the {\it Sf\_glx} project, with a very low number of exposures per sky
           position. Galaxies with no reported central fluxes have no identifiable
           central concentration: the radial surface brightness profile is
           consistent with a disk alone at our angular resolution (NGC\,4580 rather
           shows a smooth central plateau and NGC\,4634 is seen edge-on).
           For NGC\,7552, we used only the maps with a $3\arcsec$ pixel size, because
           those at $6\arcsec$ are strongly saturated in both filters; for the other
           galaxies mapped with both pixel sizes, we used the $6\arcsec$ sampling
           because of the higher signal to noise ratio and the more reasonable
           field of view.}
\label{tab:tab_photom}
\begin{minipage}[t]{18cm}
\begin{flushleft}
\begin{tabular}{|lr@{$\,$}lr@{$\,$}lrrr@{$\,$}lr@{$\,$}lr@{$\,$}lr@{$\,$}l|}
\hline
\noalign{\smallskip}
name & \multicolumn{2}{c}{$F_{15\, \rm tot}$} &
\multicolumn{2}{c}{$F_{7\, \rm tot}$} &\multicolumn{2}{c}{$D_{\, \rm CNR}$} &
\multicolumn{2}{c}{$F_{15\, \rm CNR}$} & \multicolumn{2}{c}{$F_{7\, \rm CNR}$} &
\multicolumn{2}{c}{$b_{15}$} & \multicolumn{2}{c|}{$b_7$} \\
~ & \multicolumn{4}{c}{(mJy)$^a$} & ($\arcsec$) & (kpc) &
\multicolumn{4}{c}{(mJy)$^a$} & \multicolumn{4}{c|}{($\mu$Jy\,arcsec$^{-2}$)} \\
\noalign{\smallskip}
\hline
\noalign{\smallskip}
  N289                     &   327.8 $\pm$ &  25.6 &   342.9$\pm$ &  14.7 & 12.9 & 1.21 &   68.5$\pm$ &   4.2 &   37.5$\pm$ &   1.8 &  588.$\pm$ &  6. & 115.$\pm$ &  4. \\
  N337                     &   297.9 $\pm$ &  24.0 &   336.1$\pm$ &  17.9 & 11.3 & 1.13 &   31.8$\pm$ &   2.2 &   28.5$\pm$ &   1.1 &  679.$\pm$ &  6. &  99.$\pm$ &  4. \\
  N613                     &  1566.5 $\pm$ & 104.0 &  1473.3$\pm$ &  71.4 & 19.1 & 1.62 &  557.1$\pm$ &  47.8 &  353.4$\pm$ &  31.3 &  534.$\pm$ &  6. & 101.$\pm$ &  4. \\
 N1022                     &   802.3 $\pm$ &  86.4 &   444.4$\pm$ &  45.3 & 15.0 & 1.34 &  748.9$\pm$ &  81.2 &  350.8$\pm$ &  39.2 &  744.$\pm$ &  9. & 139.$\pm$ &  4. \\
 N1097                     &  2269.2 $\pm$ & 167.4 &  2128.6$\pm$ & 125.4 & 45.6 & 3.21 & 1730.3$\pm$ &  92.6 & 1285.1$\pm$ &  65.2 &  416.$\pm$ &  6. &  74.$\pm$ &  4. \\
 N1365                     &  4436.7 $\pm$ & 764.5 &  3691.9$\pm$ & 616.6 & 42.6 & 3.49 & 3163.2$\pm$ & 420.2 & 2019.0$\pm$ & 301.2 &  370.$\pm$ &  4. &  72.$\pm$ &  2. \\
 N1433                     &   355.3 $\pm$ &  41.0 &   381.3$\pm$ &  33.8 & 31.1 & 1.75 &  117.2$\pm$ &   3.3 &  105.3$\pm$ &   2.4 &  352.$\pm$ &  5. &  61.$\pm$ &  4. \\
 N1530                     &   606.1 $\pm$ &  39.2 &   573.9$\pm$ &  39.1 & 27.7 & 4.92 &  267.6$\pm$ &  10.2 &  217.0$\pm$ &   9.8 &  345.$\pm$ &  5. &  61.$\pm$ &  4. \\
 N1672                     &  2020.5 $\pm$ & 123.0 &  1985.0$\pm$ & 129.2 & 32.4 & 2.28 & 1179.8$\pm$ &  74.7 &  956.7$\pm$ &  69.8 &  340.$\pm$ &  5. &  66.$\pm$ &  4. \\
 N4027                     &   676.7 $\pm$ &  95.5 &   775.8$\pm$ &  68.2 & 10.5 & 1.30 &   31.7$\pm$ &   6.4 &   32.4$\pm$ &   4.3 &  900.$\pm$ &  4. & 171.$\pm$ &  2. \\
 N4535                     &  1127.9 $\pm$ & 181.4 &  1136.6$\pm$ &  68.9 & 23.2 & 1.89 &  153.4$\pm$ &  15.8 &  111.8$\pm$ &  14.2 & 1043.$\pm$ &  5. & 170.$\pm$ &  2. \\
 N4691                     &   795.9 $\pm$ & 185.6 &   613.5$\pm$ &  83.1 & 44.9 & 4.90 &  730.5$\pm$ &  81.4 &  510.9$\pm$ &  53.5 & 1374.$\pm$ &  5. & 237.$\pm$ &  2. \\
 N4736 {\scriptsize ($-$)} &  4204.5 $\pm$ & 240.6 &  3913.9$\pm$ & 225.8 & 21.6 & 0.45 &  566.2$\pm$ &  39.2 &  540.6$\pm$ &  49.0 &  473.$\pm$ & 10. &  97.$\pm$ &  3. \\
 N5194                     &  8003.2 $\pm$ & 493.5 &  8598.7$\pm$ & 552.1 & 88.9 & 3.32 & 2032.7$\pm$ &  33.4 & 1869.3$\pm$ &  40.5 &  412.$\pm$ & 10. &  74.$\pm$ & 12. \\
 N5236 \#                  & 20098.4 $\pm$ & 803.7 & 18474.9$\pm$ & 899.7 & 36.8 & 0.84 & 3473.9$\pm$ & 200.2 & 2656.4$\pm$ & 203.3 & 1096.$\pm$ &  5. & 233.$\pm$ &  3. \\
 N5383                     &   332.6 $\pm$ &  61.9 &   350.2$\pm$ &  62.1 & 32.1 & 5.89 &  185.8$\pm$ &  20.1 &  171.3$\pm$ &  20.9 &  394.$\pm$ &  3. &  68.$\pm$ &  2. \\
 N5457                     &  5424.3 $\pm$ & 322.0 &  6034.0$\pm$ & 116.7 & 35.2 & 0.92 &  129.0$\pm$ &   9.2 &  114.5$\pm$ &   4.7 &  361.$\pm$ &  2. &  62.$\pm$ &  1. \\
 N6744 {\scriptsize ($-$)} &  1497.4 $\pm$ & 125.7 &  2419.4$\pm$ &  52.3 & 32.4 & 1.64 &   26.6$\pm$ &   5.6 &   50.3$\pm$ &   2.1 &  485.$\pm$ &  6. & 103.$\pm$ &  3. \\
 N7552 \#                  &  2767.6 $\pm$ & 193.7 &  1826.2$\pm$ & 168.5 & 21.3 & 2.01 & 2292.1$\pm$ & 153.3 & 1251.8$\pm$ & 137.4 &  565.$\pm$ &  7. &  94.$\pm$ &  4. \\
\multicolumn{15}{|l|}{~} \\
\multicolumn{15}{|l|}{\it Virgo cluster sample:} \\
\multicolumn{15}{|l|}{~} \\
 N4178                    &   181.5 $\pm$ &  48.0 &   228.5$\pm$ &  24.6 & 23.8 & 1.94 &   34.6$\pm$ &   4.7 &   33.4$\pm$ &   2.1 & 1051.$\pm$ &  3. & 178.$\pm$ &  2. \\
 N4192                    &   630.0 $\pm$ &  99.6 &   900.8$\pm$ &  68.3 & 29.3 & 2.39 &  132.3$\pm$ &  16.4 &  120.0$\pm$ &  23.8 &  638.$\pm$ &  3. &  96.$\pm$ &  2. \\
 N4293 {\scriptsize (+)}  &   188.6 $\pm$ &  42.8 &   159.5$\pm$ &  25.3 & 13.9 & 1.15 &  128.0$\pm$ &  34.9 &   50.2$\pm$ &  12.4 &  916.$\pm$ &  3. & 170.$\pm$ &  2. \\
 N4351                    &    45.6 $\pm$ &  26.3 &    52.6$\pm$ &   8.7 & 16.8 & 1.37 &   14.3$\pm$ &   4.7 &   14.8$\pm$ &   1.0 &  998.$\pm$ &  3. & 161.$\pm$ &  3. \\
 N4388 \#                 &  1008.2 $\pm$ & 244.0 &   499.4$\pm$ &  77.8 & 20.8 & 1.70 &  763.3$\pm$ & 244.9 &  224.7$\pm$ &  66.6 &  994.$\pm$ &  3. & 159.$\pm$ &  2. \\
 N4394                    &   139.0 $\pm$ &  41.0 &   161.2$\pm$ &  19.1 & 18.6 & 1.52 &   19.7$\pm$ &   2.9 &   21.6$\pm$ &   3.2 &  905.$\pm$ &  3. & 165.$\pm$ &  2. \\
 N4413                    &    93.0 $\pm$ &  31.4 &    89.3$\pm$ &  11.0 & 14.2 & 1.16 &   27.0$\pm$ &   4.0 &   22.4$\pm$ &   2.6 &  992.$\pm$ &  3. & 158.$\pm$ &  3. \\
 N4430                    &    98.0 $\pm$ &  23.5 &   132.5$\pm$ &  13.9 & ~    & ~    & ~           & ~     & ~           & ~     &  986.$\pm$ &  4. & 166.$\pm$ &  3. \\
 N4438 {\scriptsize (+)}  &   209.1 $\pm$ &  34.7 &   231.9$\pm$ &  26.9 & 21.0 & 1.71 &  123.8$\pm$ &  22.7 &  103.3$\pm$ &  10.0 &  906.$\pm$ &  3. & 178.$\pm$ &  2. \\
 N4450 {\scriptsize (+)}  &   169.7 $\pm$ &  42.5 &   185.1$\pm$ &  14.6 & 17.3 & 1.41 &   28.5$\pm$ &   4.5 &   28.1$\pm$ &   2.9 &  879.$\pm$ &  3. & 161.$\pm$ &  2. \\
 N4491                    &    81.1 $\pm$ &  25.2 &    30.5$\pm$ &   7.6 & 10.4 & 0.85 &   73.4$\pm$ &  21.4 &   18.3$\pm$ &   4.0 & 1052.$\pm$ &  3. & 169.$\pm$ &  3. \\
 N4498                    &    94.6 $\pm$ &  19.2 &   112.9$\pm$ &  11.8 & 14.7 & 1.20 &   11.6$\pm$ &   2.2 &   11.2$\pm$ &   1.0 &  876.$\pm$ &  3. & 160.$\pm$ &  2. \\
 N4506                    &    12.7 $\pm$ &   5.2 &    21.1$\pm$ &   9.8 & 13.0 & 1.06 &    7.5$\pm$ &   3.4 &    7.6$\pm$ &   0.9 &  895.$\pm$ &  3. & 149.$\pm$ &  3. \\
 N4567 \dag               &   293.4 $\pm$ &  15.5 &   317.9$\pm$ &  16.4 & 18.0 & 1.47 &   50.0$\pm$ &   5.7 &   41.3$\pm$ &   2.3 &  979.$\pm$ &  3. & 174.$\pm$ &  2. \\
 N4568 \dag               &  1099.0 $\pm$ & 127.6 &  1074.7$\pm$ &  64.8 & 17.5 & 1.42 &  230.2$\pm$ &  47.5 &  172.2$\pm$ &  23.0 &  979.$\pm$ &  3. & 174.$\pm$ &  2. \\
 N4569                    &   939.3 $\pm$ & 125.1 &   843.5$\pm$ &  54.1 & 21.3 & 1.73 &  289.2$\pm$ &  88.1 &  144.1$\pm$ &  23.4 &  871.$\pm$ &  3. & 144.$\pm$ &  2. \\
 N4579                    &   619.2 $\pm$ &  85.1 &   672.5$\pm$ &  37.5 & 26.4 & 2.15 &  152.2$\pm$ &  36.4 &   96.4$\pm$ &   9.7 &  973.$\pm$ &  3. & 170.$\pm$ &  2. \\
 N4580                    &   103.9 $\pm$ &  24.2 &   102.6$\pm$ &   7.7 & ~    & ~    & ~           & ~     & ~           & ~     & 1006.$\pm$ &  3. & 155.$\pm$ &  3. \\
 N4633                    &    30.0 $\pm$ &   9.5 &    30.3$\pm$ &   9.1 & ~    & ~    & ~           & ~     & ~           & ~     &  830.$\pm$ &  3. & 139.$\pm$ &  2. \\
 N4634                    &   258.2 $\pm$ &  40.7 &   278.3$\pm$ &  35.0 & ~    & ~    & ~           & ~     & ~           & ~     &  830.$\pm$ &  3. & 139.$\pm$ &  2. \\
 N4647                    &   472.3 $\pm$ &  32.0 &   474.3$\pm$ &  17.2 & 16.9 & 1.38 &   61.1$\pm$ &   6.9 &   52.8$\pm$ &   2.7 &  849.$\pm$ &  3. & 183.$\pm$ &  2. \\
 N4654                    &  1018.6 $\pm$ &  78.4 &  1049.4$\pm$ &  42.9 & 15.5 & 1.26 &   92.5$\pm$ &  25.9 &   82.9$\pm$ &  11.2 &  823.$\pm$ &  3. & 174.$\pm$ &  2. \\
 N4689                    &   329.7 $\pm$ &  37.4 &   340.9$\pm$ &  16.3 & ~    & ~    & ~           & ~     & ~           & ~     &  796.$\pm$ &  3. & 132.$\pm$ &  2. \\
\noalign{\smallskip}
\hline
\end{tabular}
\end{flushleft}
\end{minipage}
\end{table*}
\setcounter{table}{1}
\begin{table*}[!t]
\caption[]{continued.}
\begin{minipage}[t]{18cm}
\begin{flushleft}
\begin{tabular}{|lr@{$\,$}lr@{$\,$}lrrr@{$\,$}lr@{$\,$}lr@{$\,$}lr@{$\,$}l|}
\hline
\noalign{\smallskip}
name & \multicolumn{2}{c}{$F_{15\, \rm tot}$} &
\multicolumn{2}{c}{$F_{7\, \rm tot}$} &\multicolumn{2}{c}{$D_{\, \rm CNR}$} &
\multicolumn{2}{c}{$F_{15\, \rm CNR}$} & \multicolumn{2}{c}{$F_{7\, \rm CNR}$} &
\multicolumn{2}{c}{$b_{15}$} & \multicolumn{2}{c|}{$b_7$} \\
~ & \multicolumn{4}{c}{(mJy)\footnote{The conversion from flux densities to fluxes
   is: $F {\rm (W\,m}^{-2}) = 10^{-14} F_{\lambda} {\rm (Jy)} \times \Delta_{\nu}(\lambda) {\rm (THz)}$,
   with the filter widths $\Delta_{\nu}(15) = 6.75\, {\rm THz}$
   and $\Delta_{\nu}(7) = 16.18\, {\rm THz}$. \\
   \hspace*{1ex} \#~: nucleus saturated (for NGC\,5236: both at 15 and 7\,$\mu$m,
   but more severely at 15\,$\mu$m since the same gain and integration time were
   used for both filters and since $F_{15}/F_7$ is above 1 in electronic units;
   for NGC\,7552: slightly at 15\,$\mu$m, but not at 7\,$\mu$m; for the central
   pixel of NGC\,4388: at 15\,$\mu$m, but not at 7\,$\mu$m; for NGC\,3620: at
   7\,$\mu$m but not at 15\,$\mu$m, which is possible because the integration time
   was respectively 5\,s and 2\,s; for NGC\,4102 and 6946: both at 7 and 15\,$\mu$m,
   but more severely at 7\,$\mu$m, with the same configuration as NGC\,3620). Thus,
   $F_{15}/F_7$ colors in central regions are respectively lower limits for
   NGC\,5236, 7552 and 4388 and upper limits for NGC\,3620, 4102 and 6946. \\
   \hspace*{1ex} {\scriptsize ($-$)}~: The field of view is too small to allow a
   precise determination of the backgroud level and total fluxes are lower limits.
   The error bars are only formal. The comparison of our measurements with those of
   \citet{Rice} at 12\,$\mu$m, inside the IRAS band 8--15\,$\mu$m which overlaps
   with our 5--8.5\,$\mu$m and 12--18\,$\mu$m bands, indicates that we miss of the
   order of 15\% of total fluxes for NGC\,4736 and between 15 and 45\% for NGC\,6744,
   provided IRAS fluxes are not overestimated as this is often the case for
   co-added observations. \\
   \hspace*{1ex} {\scriptsize (+)}~: From their spectral energy distributions
   shown by \citet{Boselli}, these galaxies probably have a non-negligible
   contribution from the Rayleigh-Jeans tail of cold stars to their 7\,$\mu$m
   emission. We did not attempt to remove this contribution, because it would
   require a careful modelling of stellar populations. \\
   \hspace*{1ex} \dag~: The disks of these galaxies slightly overlap in projection.
   We attempted to separate them by the means of a mask defined visually, but the
   disk fluxes are much more uncertain than estimated.}} &
($\arcsec$) & (kpc) & \multicolumn{4}{c}{(mJy)$^a$} &
\multicolumn{4}{c|}{($\mu$Jy\,arcsec$^{-2}$)} \\
\noalign{\smallskip}
\hline
\noalign{\smallskip}
\multicolumn{15}{|l|}{\it Supplementary galaxies:} \\
\multicolumn{15}{|l|}{~} \\
  N986    &  1050.0 $\pm$ &   79.0 &   801.5$\pm$ &  12.1 & 22.9 & 2.57 &  657.0$\pm$ &  67.1 & 316.1$\pm$ &  11.5 &  372.$\pm$ & 4. &  63.$\pm$ & 2. \\
 N1326    &   287.8 $\pm$ &   58.8 &   284.9$\pm$ &  18.8 & 33.1 & 2.71 &  258.4$\pm$ &  35.6 & 212.3$\pm$ &  16.1 &  353.$\pm$ & 4. &  62.$\pm$ & 2. \\
 N1385    &   782.3 $\pm$ &   62.7 &   815.7$\pm$ &  27.1 & 12.9 & 1.10 &   79.9$\pm$ &   7.2 &  68.5$\pm$ &   4.6 &  385.$\pm$ & 4. &  73.$\pm$ & 2. \\
 N3583    &   448.4 $\pm$ &   52.5 &   425.6$\pm$ &  19.1 & 14.0 & 2.31 &   87.2$\pm$ &   4.2 &  70.0$\pm$ &   5.0 &  517.$\pm$ & 5. &  97.$\pm$ & 2. \\
 N3620 \# &  1199.8 $\pm$ &  408.6 &   723.1$\pm$ &  89.2 & 30.2 & 3.47 & 1087.2$\pm$ & 408.3 & 537.2$\pm$ &  74.7 &  386.$\pm$ & 4. &  81.$\pm$ & 2. \\
 N3683    &   755.5 $\pm$ &   71.3 &   791.1$\pm$ &  57.4 & ~	 & ~	& ~	      & ~     & ~          & ~     &  452.$\pm$ & 5. &  88.$\pm$ & 2. \\
 N3705    &   307.4 $\pm$ &   73.6 &   348.2$\pm$ &  32.6 & 14.1 & 1.16 &   28.6$\pm$ &   1.8 &  28.9$\pm$ &   2.1 &  686.$\pm$ & 6. &  96.$\pm$ & 2. \\
 N3885    &   396.0 $\pm$ &   16.0 &   342.9$\pm$ &  10.2 & 30.7 & 4.14 &  363.9$\pm$ &   8.9 & 293.0$\pm$ &   9.8 &  738.$\pm$ & 5. & 150.$\pm$ & 2. \\
 N4041    &   751.9 $\pm$ &  128.1 &   792.8$\pm$ & 164.5 & 36.4 & 4.01 &  482.8$\pm$ &  69.5 & 476.7$\pm$ & 123.7 &  428.$\pm$ & 6. &  96.$\pm$ & 4. \\
 N4102 \# &  1712.9 $\pm$ &  561.0 &   808.8$\pm$ & 125.9 & 24.3 & 2.00 & 1419.2$\pm$ & 570.7 & 462.3$\pm$ & 113.7 &  443.$\pm$ & 5. &  81.$\pm$ & 2. \\
 N4519    &   233.8 $\pm$ &   67.6 &   177.8$\pm$ &  18.3 & 11.2 & 0.91 &   85.3$\pm$ &  24.3 &  20.3$\pm$ &   0.8 & 1050.$\pm$ & 6. & 183.$\pm$ & 2. \\
 N4713    &   209.4 $\pm$ &   55.7 &   223.6$\pm$ &  13.3 & ~	 & ~	& ~	      & ~     & ~          & ~     &  845.$\pm$ & 6. & 131.$\pm$ & 2. \\
 N5430    &   530.1 $\pm$ &  136.1 &   364.8$\pm$ & 153.7 & 20.7 & 4.05 &  273.0$\pm$ &  80.3 & 152.5$\pm$ & 106.9 &  375.$\pm$ & 5. &  81.$\pm$ & 4. \\
 N5786    &   380.2 $\pm$ &   81.6 &   345.8$\pm$ &  25.5 & 19.3 & 3.72 &  120.1$\pm$ &  15.5 &  85.7$\pm$ &   5.4 &  823.$\pm$ & 7. & 173.$\pm$ & 3. \\
 N5937    &   616.1 $\pm$ &  128.9 &   562.3$\pm$ &  94.6 &  7.4 & 1.33 &  157.7$\pm$ &  90.1 &  85.2$\pm$ &  29.1 & 1140.$\pm$ & 8. & 234.$\pm$ & 4. \\
 N5962    &   508.8 $\pm$ &   37.2 &   485.5$\pm$ &  17.0 & ~	 & ~	& ~	      & ~     & ~          & ~     &  499.$\pm$ & 5. &  97.$\pm$ & 2. \\
 N6156    &   827.0 $\pm$ &  149.2 &   697.9$\pm$ & 118.0 & ~	 & ~	& ~	      & ~     & ~          & ~     &  457.$\pm$ & 6. &  86.$\pm$ & 5. \\
 N6753    &   646.8 $\pm$ &   60.1 &   586.4$\pm$ &  16.9 & 32.3 & 6.64 &  247.8$\pm$ &   7.3 & 218.7$\pm$ &   3.1 &  466.$\pm$ & 5. &  78.$\pm$ & 2. \\
 N6824    &   394.6 $\pm$ &   98.1 &   408.3$\pm$ & 101.3 &  9.8 & 2.11 &   81.5$\pm$ &  30.3 &  74.4$\pm$ &  23.6 &  335.$\pm$ & 5. &  69.$\pm$ & 4. \\
 N6946 \# & 10651.6 $\pm$ & 1767.2 & 11648.8$\pm$ & 678.6 & 30.2 & 0.81 & 1730.4$\pm$ & 747.0 & 970.6$\pm$ & 277.5 &  424.$\pm$ & 5. &  98.$\pm$ & 2. \\
 N7218    &   273.5 $\pm$ &   54.1 &   260.6$\pm$ &  12.8 & ~	 & ~	& ~	      & ~     & ~          & ~     & 1013.$\pm$ & 7. & 162.$\pm$ & 2. \\
 N7418    &   455.5 $\pm$ &  113.3 &   469.8$\pm$ &  32.9 & 17.9 & 1.55 &   38.6$\pm$ &   3.1 &  31.1$\pm$ &   1.2 &  640.$\pm$ & 5. & 112.$\pm$ & 2. \\
 N7771    &   615.2 $\pm$ &  116.6 &   526.8$\pm$ &  57.4 & 19.4 & 5.41 &  323.8$\pm$ &  76.1 & 247.1$\pm$ &  33.9 &  869.$\pm$ & 6. & 180.$\pm$ & 3. \\
 I1953    &   222.9 $\pm$ &   37.0 &   186.2$\pm$ &  13.7 & 14.3 & 1.53 &  116.2$\pm$ &  23.1 &  34.4$\pm$ &   2.8 &  429.$\pm$ & 3. &  85.$\pm$ & 2. \\
 I2554    &   887.6 $\pm$ &  345.2 &   733.5$\pm$ & 242.6 & 17.9 & 1.45 &  503.9$\pm$ & 369.7 & 264.0$\pm$ & 183.4 &  376.$\pm$ & 5. &  81.$\pm$ & 4. \\
 I5325    &   373.9 $\pm$ &   42.2 &   364.6$\pm$ &  17.4 & ~	 & ~	& ~	      & ~     & ~          & ~     &  537.$\pm$ & 5. &  96.$\pm$ & 2. \\
ESO317    &   287.2 $\pm$ &   59.2 &   239.8$\pm$ &  38.4 & 19.9 & 3.17 &  222.2$\pm$ &  32.4 & 162.1$\pm$ &  20.7 &  519.$\pm$ & 5. & 112.$\pm$ & 3. \\
\noalign{\smallskip}
\hline
\end{tabular}
\end{flushleft}
\end{minipage}
\end{table*}